\documentclass[pdflatex,sn-mathphys-num]{sn-jnl}
\usepackage{graphicx}
\usepackage{xcolor}
\usepackage{amsmath,amssymb,amsfonts}
\usepackage{amsthm}
\usepackage{bm}
\usepackage{mathrsfs}
\usepackage{multirow}
\usepackage{booktabs}
\usepackage{ragged2e}
\usepackage{subcaption}
\usepackage{hyperref}
\usepackage{algorithm}
\usepackage{algorithmicx}
\usepackage{algpseudocode}
\usepackage{listings}

\begin{document}

\title[Article Title]{Effect of Colored Noise on Coupled Thermoacoustic Oscillators}

\author[1]{\fnm{Robert} \sur{Rai}}

\author[1]{\fnm{Yuvrajsingh} \sur{Patil}}

\author[2]{\fnm{Lipika} \sur{Kabiraj}}

\author[3]{\fnm{Aditya} \sur{Saurabh}}

\author*[1]{\fnm{Chandrakala} \sur{Meena}}\email{chandrakala@iiserpune.ac.in}

\affil[1]{\orgdiv{Department of Physics},
\orgname{Indian Institute of Science Education and Research Pune},
\orgaddress{\postcode{411008}, \state{Maharashtra}, \country{India}}}

\affil[2]{\orgdiv{Department of Mechanical Engineering},
\orgname{Indian Institute of Technology Ropar},
\orgaddress{\postcode{140001}, \state{Punjab}, \country{India}}}

\affil[3]{\orgdiv{Department of Mechanical Engineering},
\orgname{Indian Institute of Technology Kanpur},
\orgaddress{\postcode{208016}, \state{Uttar Pradesh}, \country{India}}}
\abstract{
Noise can significantly influence thermoacoustic dynamics, yet the role of noise color in coupled thermoacoustic oscillators remains largely unexplored. Here, we examine the influence of colored noise on the dynamics of coupled thermoacoustic systems. The system consists of two coupled Rijke tube oscillators with time-delay and dissipative coupling. Stochastic forcing is modeled as an additive Ornstein–Uhlenbeck (OU) process, such that white and colored noise contain equal power within a band around the system's natural frequency. We find that noise influences the system most prominently near the transition between limit-cycle oscillations (LCO) and amplitude death (AD) states, where increasing noise amplitude smoothen the transition and reduces the extent of the AD regions. Our analysis reveals the emergence of coherence resonance near instability threshold under both white and colored noise. The peak coherence factor varies with the noise color, with the largest peak coherence observed for colored noise whose correlation time is much shorter than the acoustic time scale. White noise and the shortest-correlated OU noise exert the strongest influence on both the pressure amplitude response and the coherence resonance. Overall, our results show that, under both coupling mechanisms, colored noise induces qualitatively similar trends in the system response, governed by its amplitude and correlation time.}
\keywords{Thermoacoustic Instability, Stochastic Dynamics, Coupled Oscillators, Amplitude Death, Coherence Resonance, Colored Noise}
\maketitle
\section{Introduction}\label{section_1}
Practical combustors are prone to severe vibrations, structural damage, increased emissions, and, in extreme cases, catastrophic engine failure due to a phenomenon known as thermoacoustic instability \cite{lieuwen2005combustion,balasubramanian2008thermoacoustic,sujith2021thermoacoustic}. This instability arises from the coupling between acoustic pressure fluctuations and unsteady heat release, forming a nonlinear feedback mechanism that leads to the growth of self-sustained oscillations, which eventually saturate into limit-cycle oscillations (LCO) \cite{lieuwen2002experimental,morgans2025thermoacoustic}. The transition to such oscillatory states is typically associated with a Hopf bifurcation, and in many thermoacoustic configurations, it exhibits subcritical behavior characterized by bistability and hysteresis \cite{matveev2003thermoacoustic,subramanian2013subcritical,singh2025synchronization}. As a result, thermoacoustic systems are highly sensitive to perturbations, particularly near stability boundaries where small disturbances can trigger large changes in system dynamics. Understanding how intrinsic acoustic modes interact with external perturbations is therefore central to predicting and controlling thermoacoustic behavior.

Low-order models, particularly the Rijke tube configuration, have served as canonical platforms for investigating the nonlinear dynamics of thermoacoustic systems \cite{balasubramanian2008thermoacoustic,nair2015reduced,liao2024low}. These models capture key features such as Hopf bifurcations, hysteresis, bistability, and limit-cycle oscillations, while remaining analytically and computationally tractable. Beyond single-oscillator dynamics, coupling between thermoacoustic elements has received considerable attention as a mechanism for oscillation suppression \cite{sharma2014effect,juniper2018sensitivity,gotoda2014detection,poinsot2017prediction,kobayashi2019early}. When multiple thermoacoustic oscillators interact, their collective behavior can differ significantly from that of isolated systems, giving rise to synchronization, phase locking, and oscillation suppression \cite{guan2022synchronization,vorgias2025experimental}. The study of amplitude death (AD) is of particular interest in thermoacoustic systems because it provides a mechanism in which coupling stabilizes a previously unstable steady state and suppresses oscillations completely \cite{biwa2015amplitude,meena2017effect,thomas2018effect}. In thermoacoustic systems modeled using coupled Rijke tubes, it has been shown that time-delay coupling can induce amplitude suppression even in identical oscillators \cite{hyodo2018stabilization,hyodo2020suppression}, while dissipative coupling requires sufficient detuning between their natural frequencies \cite{delage2018bifurcation,hyodo2019amplitude}. The combined action of time-delay and dissipative coupling enlarges the parameter space supporting amplitude suppression, enhancing the robustness of coupling-based control \cite{thomas2018effect, zheng2025amplitude}. However, in nonidentical coupled thermoacoustic oscillators, parameter mismatch, such as differences in heater power, reduces the extent of AD regions and can lead to asymmetric behavior, where one oscillator exhibits amplification while the other is suppressed \cite{doranehgard2022quenching}.

Despite these advances, most investigations of coupling-induced suppression have been carried out under deterministic assumptions. In practice, thermoacoustic systems operate in turbulent environments where stochastic forcing is unavoidable \cite{kabiraj2019review}. Turbulent heat release, vortex shedding, and hydrodynamic instabilities introduce broadband fluctuations that continuously perturb the acoustic field \cite{sujith2020complex,sahay2025vortical}. These fluctuations interact nonlinearly with the system dynamics, especially near bifurcation points where stability margins are reduced \cite{vishnoi2024effect}. 
In coupled thermoacoustic oscillators, such stochastic effects modify both the onset and the regime of amplitude suppression \cite{thomas2018noise}.

Previous investigations of stochastic effects in thermoacoustic oscillators have largely assumed additive white noise forcing \cite{noiray2013deterministic,thomas2018noise}. While this assumption simplifies analysis, it does not accurately represent combustion noise observed in experiments, which exhibits finite temporal correlation and a non-uniform spectral distribution with a predominantly low-pass character \cite{rajaram2009acoustic,nawroth2013flow}. The presence of temporal correlation introduces memory into the stochastic forcing and redistributes energy toward lower frequencies, often close to the dominant acoustic mode \cite{BUSCH2003}. As a result, the interaction between noise and system dynamics cannot be fully captured using white-noise models. The inclusion of finite correlation time raises an important question regarding its role in coupled thermoacoustic systems. Since amplitude suppression is governed by the interaction between coupling mechanisms and the least stable acoustic mode, temporally correlated fluctuations may influence this interaction in nontrivial ways. Reduced high-frequency content may limit broadband excitation, whereas persistent correlated fluctuations can enhance coherent amplification near marginal stability.

Recent studies on thermoacoustic systems subjected to colored noise have begun to address the role of temporal correlation in modifying system dynamics, particularly near instability thresholds \cite{li2020stochastic, li2023effects,vishnoi2024effect,zhang2025tipping}. These studies suggest that while noise amplitude plays a dominant role in determining the overall response, temporal correlation can influence the nature of fluctuations and the spectral characteristics of the system. However, the effect of colored noise on coupled thermoacoustic oscillators, especially in the context of amplitude suppression and near instability thresholds, remains largely unexplored. In addition to modifying bifurcation characteristics, stochastic forcing can also induce organization in nonlinear systems through coherence resonance (CR), where oscillations become most regular at an optimal level of noise. Although first identified in excitable systems, CR has since been observed in chaotic oscillators and coupled nonlinear systems \cite{liu2001coherence,zhan2002coherence}, where noise can give rise to coherent oscillations even when deterministic oscillations are absent. Similar effects have been reported in systems exhibiting noise-induced intermittency, where the regularity of the response is governed by the interaction between deterministic dynamics and stochastic perturbations \cite{neiman1997coherence}. In spatially extended systems, such as oscillator arrays and directionally coupled networks, noise has been shown to promote synchronization and generate coherent spatiotemporal patterns \cite{okano2007array,werner2011coherence}. In systems that support localized states, stochastic forcing can destabilize these states and drive transitions toward more uniform configurations, reflecting the sensitivity of such attractors to perturbations \cite{balachandran2022dynamics}. When the forcing is temporally correlated, its impact becomes dependent on how the noise time scale compares with the system dynamics, and studies have shown that colored noise can either enhance synchronization or weaken resonance \cite{ma2008coherence,liu2025optimizing}. Coherence resonance has also been observed in thermoacoustic systems, where noise can induce coherent oscillations in a linearly stable regime prior to the Hopf bifurcation \cite{kabiraj2015coherence,vishnoi2024effect}.

In this work, we investigate the influence of colored noise on two coupled thermoacoustic oscillators modeled as horizontal Rijke tubes. The oscillators interact through both time-delay and dissipative coupling mechanisms, enabling a systematic comparison of two distinct coupling configurations. The stochastic forcing is modeled using an Ornstein--Uhlenbeck
(OU) process \cite{bonciolini2017output, vishnoi2024gasturbine}, which captures the low-pass spectral characteristics of combustion noise. By independently varying the noise amplitude and correlation time using a power-normalized OU process, we examine their individual effects on coupled thermoacoustic dynamics. In particular, we assess how colored stochastic forcing modifies amplitude suppression and the extent of the AD region under both time-delay and dissipative coupling using one and two-parameter response analyses. We also investigate coherence resonance through the coherence factor to quantify the influence of noise correlation time on spectral coherence near instability thresholds. This combined analysis provides a unified assessment of how the amplitude and temporal correlation of stochastic forcing influence emergent dynamics and coherence resonence in coupled thermoacoustic oscillators.

This paper is organized as follows: Section \ref{section_2} presents methodology that includes the low-order modeling framework for the coupled thermoacoustic oscillators (Sec. \ref{section_2a}, \ref{section_2b}) and the formulation of the colored noise (Sec. \ref{section_2c}). The results are discussed in Section \ref{section_3}, first with the time-delay coupled system (Sec. \ref{section_3A}), and followed by the dissipatively coupled system (Sec. \ref{section_3B}). Finally, Section \ref{section_4} summarizes the key findings and conclusions of the study.
\section{Methodology}\label{section_2}
\subsection{Rijke Tube Model}\label{section_2a}We first consider a single Rijke tube, represented as a horizontal duct containing a compact heater that serves as a localized heat source. Under low Mach number approximation with negligible mean flow and mean temperature gradients \cite{nicoud2009zero}, the governing linearized momentum and energy equations for the acoustic field can be written in the following dimensionless form:
\begin{align}
    &\gamma M \frac{\partial U'}{\partial t} + \frac{\partial P'}{\partial x} =0 \label{momentum}\\
    &\frac{\partial P'}{\partial t} + \gamma M \frac{\partial U'}{\partial x} + \zeta P'= (\gamma -1) \dot{Q}'\delta(x-x_f)\label{energy} 
\end{align}
Here, $x$ denotes the longitudinal coordinate of the duct normalized by its length $L_a$ and $t$ represents time normalized by $L_a/c_0$,  where $c_0$ is the speed of sound; $U'(x,t)$ and $P'(x,t)$ are the non-dimensional acoustic velocity and acoustic pressure fluctuations; $\delta(x-x_f)$ is the Dirac delta function representing the compact heat source located at the normalized heater location $x_f$; $\gamma$ is the specific heat ratio and $M$ is the Mach number.
The heat release rate, $\dot{Q}'$, is modeled using the modified King's law, proposed by Heckl \cite{heckl1990non},
\begin{equation*}
    \dot{Q}' = \frac{M\gamma K}{2(\gamma -1)}\Bigg[\sqrt{\left|\frac{1}{3}+U_f'(t-t_f)\right|}-\sqrt{\frac{1}{3}}\Bigg]
    \label{king's_law}
\end{equation*}
where $K$ is the non-dimensional heater power and $U_f'$ is the acoustic velocity at $x_f$. The modified King's law introduces a nonlinear dependence of the heat-release rate on the delayed acoustic velocity, thereby rendering the resulting thermoacoustic system nonlinear. Eqs. \ref{momentum} and \ref{energy} are solved by using the Galerkin technique that transforms these partial differential equations into a set of discrete ordinary differential equations \cite{balasubramanian2008thermoacoustic}. We express $U'$ and $P'$ as orthogonal bases that match the open-open boundary conditions.
\begin{align}
    U' = \sum_{j=1}^{N}\eta_j (t) cos(j\pi x), \quad \quad\ 
    P' = -\sum_{j=1}^{N} \dot{\eta} (t)_j\frac{\gamma M}{j \pi} sin(j \pi x)
    \label{galerkin}
\end{align}
where $\eta_j (t)$ and $\dot{\eta_j} (t)$ are the time-dependent amplitudes of the acoustic velocity $U'$ and acoustic pressure $P'$ for the $j$th Galerkin mode, respectively.

By decomposing the Galerkin modes in Eqs. \ref{momentum} and \ref{energy}, we arrive at the set of $2N$ ODEs, 
\begin{align}
    \frac{d\eta_j}{dt} &= \dot{\eta}_j, \label{ode0} \\
    \frac{d\dot{\eta}_j}{dt}
    + 2\zeta_j \omega_j \dot{\eta}_j
    + \omega_j^2 \eta_j 
    &= -j \pi K 
    \Bigg[\sqrt{\left|\frac{1}{3}+U_f'(t-t_f)\right|}
    -\sqrt{\frac{1}{3}}\Bigg]
    \sin(j\pi x_f)
    \label{ode1}
\end{align}
Since $U_f'$ is expressed as a linear combination of the Galerkin modes (Eq. \ref{galerkin}), the heat release term in Eq. (\ref{ode1}) is nonlinear in the modal amplitudes.
Here, $\omega_j = j \pi$ is the angular frequency of the $j$th Galerkin mode, and $\zeta_j$ is the corresponding acoustic damping, which is given by:
\begin{equation}
    \zeta_j = \frac{1}{2\pi}\Bigg[C_1 \frac{\omega_j}{\omega_1} + C_2\sqrt{\frac{\omega_1}{\omega_j}} \Bigg]
\end{equation}
where $C_1$ and $C_2$ are the damping coefficients. These values are fixed at $C_1 =0.1$ and $C_2 =0.06$ following \cite{matveev2003model}. 

\begin{figure}
   \centering
\includegraphics[width=0.9\linewidth]{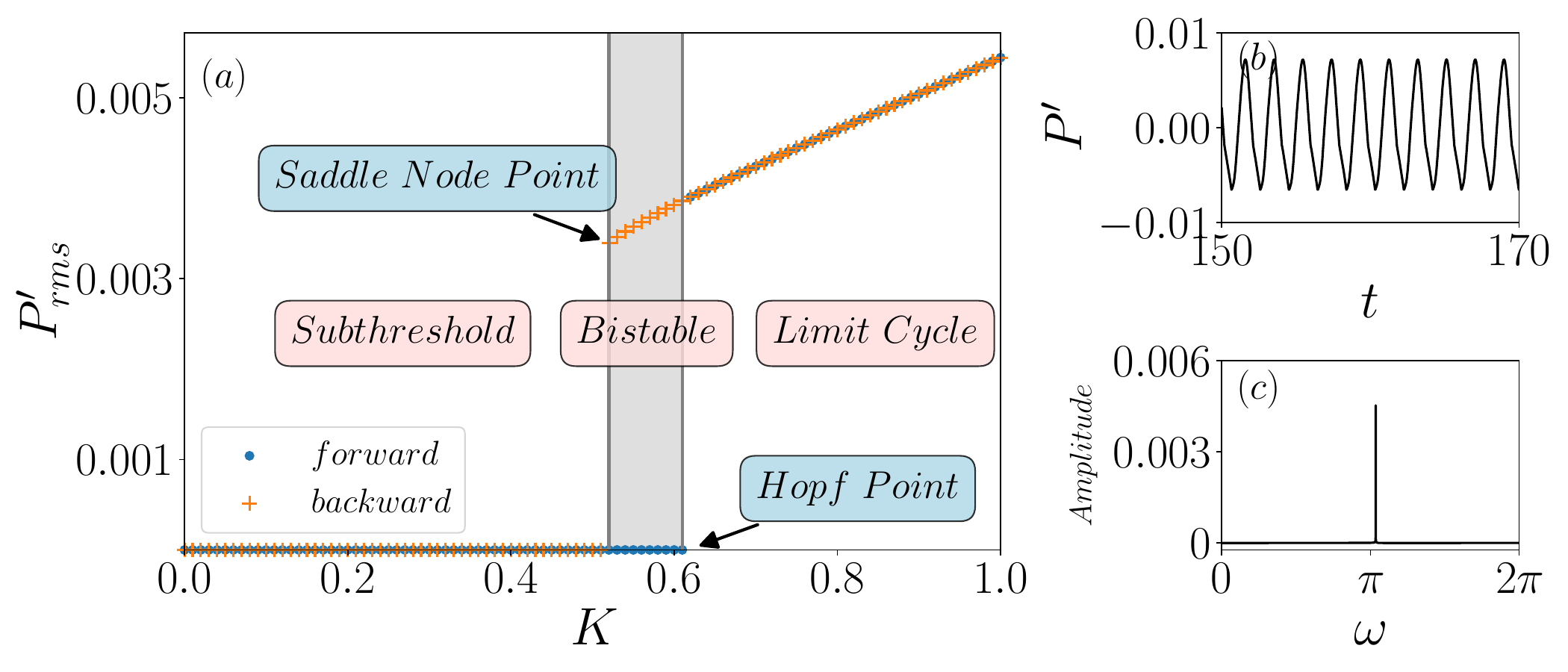}
    \caption{\justifying (a)Bifurcation diagram of a single Rijke tube showing the non-dimensional rms pressure $P_{rms}'$ as a function of the heater power $K$. The system undergoes a Hopf bifurcation at $K= 0.62$ and a saddle-node bifurcation at $K = 0.52$, indicating a bistable regime in $K \in [0.52,0.62]$. (b) Time series of the acoustic pressure and (c) amplitude spectrum of the acoustic pressure signal at $K = 0.8$. A dominant peak is observed at a non-dimensional frequency of approximately  $\omega_0 =3.25$.}
    \label{bifur}
\end{figure}

For the uncoupled Rijke-tube oscillator, the governing ODEs Eqns. (\ref{ode0},\ref{ode1}) are solved for the fixed parameters: $\gamma = 1.4, M = 0.01,t_f = 0.2,x_f = 0.25$ following \cite{thomas2018effect, vishnoi2024effect}. The detailed numerical algorithm is provided in Appendix \ref{algorithm}. We choose N = 10, which represents the total number of Galerkin modes, as additional modes give marginal contribution, as given in Appendix \ref{appendix_modes}. The bifurcation diagram shown in  Fig.~\ref{bifur}, with the heater power $K$ as the control parameter, indicates that the root mean square (RMS) pressure fluctuation $P_{rms}'$ remains zero in the subthreshold regime up to a critical value of $K \approx 0.62$ along the forward (increasing $K$) branch. Beyond this point, the system undergoes a sudden transition to a large-amplitude, self-sustained oscillatory state corresponding to the limit cycle regime. The oscillations on this high-amplitude branch are periodic in time, indicating a transition from a stable fixed point to a limit cycle via a Hopf bifurcation (Hopf point at $K=0.62$). When $K$ is decreased, the system does not immediately return to the zero amplitude state but remains on the limit-cycle branch until a saddle-node turning point at $K\approx 0.52$ is crossed, producing a hysteretic bistable interval (region between two vertical lines). These Hopf and saddle-node thresholds have good agreement with the values reported in earlier studies \cite{thomas2018effect,doranehgard2022quenching}. In the limit-cycle regime, the oscillator's dominant frequency is roughly $\omega = 2\pi f \approx 3.25$ and the acoustic time period $T \approx 2$.
\begin{figure}
    \centering
    \includegraphics[width=0.6\linewidth]{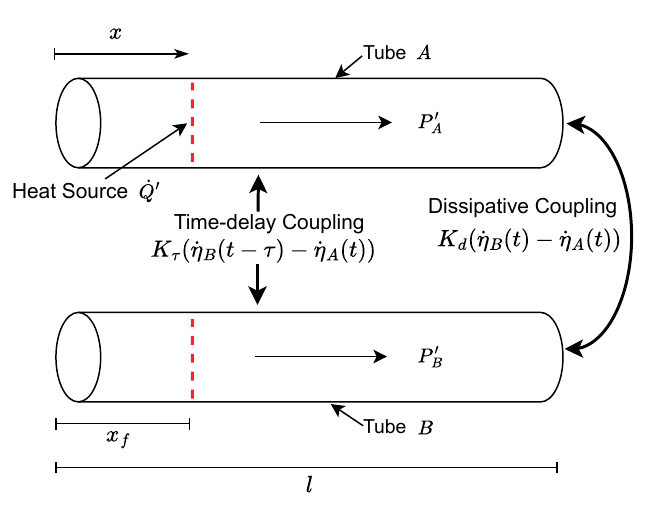}
    \caption{\justifying Schematic of two coupled Rijke tubes, each containing a heated mesh (shown as red dashed lines). A side-to-side connection between the tubes (shown using dashed lines) introduces a finite propagation delay, resulting in time-delay coupling. A head-to-head connection (shown using dashed lines) provides dissipative coupling between the oscillators.}
    \label{schematic}
\end{figure}

\subsection{Coupled Rijke Tubes}\label{section_2b}
We now consider a system of two identical Rijke tube oscillators that interact through acoustic pressure fluctuations, modeled using dissipative and time-delay coupling mechanisms. The two oscillators are distinguished using superscripts $A$ and $B$, representing the first and second tubes, respectively, as shown in Fig.~\ref{schematic}. For Oscillator $A$, the governing modal equations take the form
\begin{align}
    \frac{d\eta_j^A}{dt} &= \dot{\eta}_j^A, \label{eq1} \\
    \frac{d\dot{\eta}_j^A}{dt}
    + 2\zeta_j \omega^A_j \dot{\eta}_j^A
    + (\omega_j^A)^2 \eta_j^A 
    &= -j \pi K^A\Bigg[\sqrt{\left|\tfrac{1}{3}+U_f^{'A}(t-t_f)\right|}-\sqrt{\tfrac{1}{3}}\Bigg]\sin(j\pi x_f) \notag \\
    &+ \underbrace{K_d \big(\dot{\eta}_j^B - \dot{\eta}_j^A\big)}_{\text{Dissipative coupling}} + \underbrace{K_\tau
    \big[\dot{\eta}_j^B(t-\tau) - \dot{\eta}_j^A(t)\big]}_{\text{Time-delay coupling}}  + \underbrace{\xi^A(t).}_{\text{Noise}}
    \label{eq2}
\end{align}

The second and third terms on the right-hand side account for the dissipative and time-delay coupling contributions, respectively. The dissipative coupling represents instantaneous acoustic energy exchange and dissipation across the interconnecting junction, whereas the time-delay coupling accounts for the finite propagation time required for acoustic disturbances to travel between the oscillators through the connecting tube \cite{biwa2015amplitude,thomas2018effect,dange2019oscillation,sahay2021dynamics}. The last term $\xi(t)$ represents the stochastic forcing. Independent OU noises ($\xi^A,\xi^B$) are applied to each oscillator, with the same stochastic forcing acting on all acoustic modes within a given oscillator. The coefficients $K_d$ and $K_\tau$ denote the strengths of dissipative and time-delay coupling, respectively, while $\tau$ represents the associated delay time. Both coupling mechanisms act directly on the acoustic pressure modes $(\dot{\eta}_j)$, indicating that the interaction is global rather than spatially localized within the duct. The corresponding equations for oscillator $B$ are obtained by interchanging the superscripts $A$ and $B$ in Eq.\ref{eq2}. In the coupled configuration considered here, the parameters $K_d$, $K_\tau$, and $\tau$ are identical for both tubes. Setting $K_\tau = 0$ reduces the system to pure dissipatively coupled oscillators, whereas $K_d=0$ yields a system governed solely by time-delay interaction.

\subsection{Colored Noise Model}\label{section_2c}
Thermoacoustic oscillators operating under turbulent combustion are continuously subjected to broadband stochastic forcing. As discussed in the introduction, experimental measurements indicate that combustion noise exhibits finite temporal correlation and a predominantly low-pass spectral structure rather than ideal white-noise behavior \cite{rajaram2009acoustic, nawroth2013flow}. Accordingly, we model the stochastic forcing acting on the coupled thermoacoustic oscillators as an additive OU process \cite{bonciolini2017output, vishnoi2022system}, which represents exponentially correlated Gaussian noise with finite memory.

We first define a reference zero-mean Gaussian white noise $\epsilon_0(t)$ satisfying 
\begin{equation}
    \langle \epsilon_0(t) \epsilon_0(t') \rangle =\Gamma_0 \delta(t - t')
\end{equation}
where $\Gamma_0$ denotes the reference white-noise intensity and $\delta(t - t')$ is the Dirac delta function. The reference intensity is calibrated at $K=0$ such that the rms acoustic pressure induced by the noise is equal to the rms amplitude of the deterministic LCO at $K =0.8$ in an uncoupled Rijke tube, i.e., $P_{rms}^{noise} (K=0) = P_{rms}^{LCO} (K=0.8)$. To vary the noise amplitude, the reference white-noise is multiplied by a dimensionless noise amplitude parameter $\sigma$,
\begin{equation}
    \epsilon_\sigma (t) = \sigma \epsilon_0 (t).
\end{equation}
The corresponding noise intensity of the scaled white noise is therefore $\Gamma_\sigma = \sigma^2 \Gamma_0$ with the autocorrelation $\langle \epsilon_\sigma(t) \epsilon_\sigma(t') \rangle =\Gamma_\sigma \delta(t - t')$.

The colored noise process $\xi(t)$, independently realized for both oscillators $A$ and $B$, for the corresponding white noise $\epsilon_\sigma (t)$ is governed by the Langevin equation
\begin{equation}
    \dot{\xi}(t) = -\frac{1}{\tau_c}\xi(t) + \frac{\sqrt{D}}{\tau_c} \epsilon_\sigma (t)
\end{equation}
where $\tau_c $ denotes the non-dimensional noise correlation time, $D$ is a normalization coefficient which will be used later in the paper to adjust the power of the noise $\xi(t)$ within the frequency band of interest. The OU noise is characterized by the following mean and autocorrelation function:
\begin{equation}
    \langle \xi(t)\rangle= 0, \quad \quad \quad \quad \langle\xi(t)\xi(t')\rangle=s^2\exp\bigg(-\frac{\left| {t-t'}\right|}{\tau_c}\bigg)
\end{equation}

where $s^2$ is the variance of the process, given by $s^2 = (\Gamma_\sigma D)/(2\tau_c)$. This variance corresponds to the second moment, $\langle\xi^2\rangle$, of the OU noise.

The power spectral density (PSD) of white noise is given by 
\begin{equation}
    S_{\epsilon_\sigma \epsilon_\sigma}(\omega) = \frac{\Gamma_\sigma}{2 \, \pi}.
\end{equation} 
In the frequency domain, the OU process $\xi(t)$ can be represented as the output of a first-order linear filter driven by white noise $\epsilon (t)$, with the transfer function
\begin{equation}
    H(\omega) = \frac{\xi(\omega)}{\epsilon_\sigma(\omega)}= \frac{\sqrt{D}}{1+ i\omega \tau_c}
\end{equation}
where $i =\sqrt{-1}$ is the imaginary unit. The power spectrum of $\xi(t)$ with the underlying white noise $\epsilon_\sigma(t)$ is then given by
\begin{equation}
    S_{\xi \xi}(\omega) = \left|H(\omega)\right|^2 S_{\epsilon_\sigma \epsilon_\sigma}(\omega)=\frac{\Gamma_\sigma}{2 \, \pi} \frac{D}{1+\omega^2 \, \tau_c^2},
\end{equation}
In the limiting case $\tau_c \to 0$ and $D \to 1$, we get $S_{\xi \xi}(\omega) = S_{\epsilon_\sigma \epsilon_\sigma}(\omega)$,
indicating that the OU process converges to white noise.

To ensure a consistent comparison between white noise and colored noise across different correlation times, we generate OU noise such that the effective power provided by $\xi(t)$ and the corresponding scaled white noise $\epsilon_\sigma(t)$ within a band $[\omega_1, \omega_2]$ around the system's natural frequency $\omega_0$ remains constant:
\begin{equation}
    \int_{\omega_1}^{\omega_2} S_{\xi \xi} \, d\omega = \int_{\omega_1}^{\omega_2} S_{\epsilon_\sigma \epsilon_\sigma} \, d\omega. 
\end{equation}
Accordingly, the normalization coefficient $D$ can then be evaluated using the following expression:
\begin{equation}
    D (\tau_c) = \frac{\tau_c(\omega_2 - \omega_1)}{\tan^{-1}(\omega_2 \tau_c) - \tan^{-1} (\omega_1 \tau_c)}.
    \label{eqn17}
\end{equation}

The reference white noise is first scaled to obtain the desired noise amplitude, after which the corresponding OU process is generated using the scaled white noise as the stochastic input. The coefficient $D(\tau_c)$ ensures that the colored and white noise contain equal power within the selected band. Consequently, $\sigma$ controls the overall noise amplitude, whereas $\tau_c$ controls the temporal correlation and spectral distribution of the noise.

\begin{figure}
    \centering
    \includegraphics[width=0.6\linewidth]{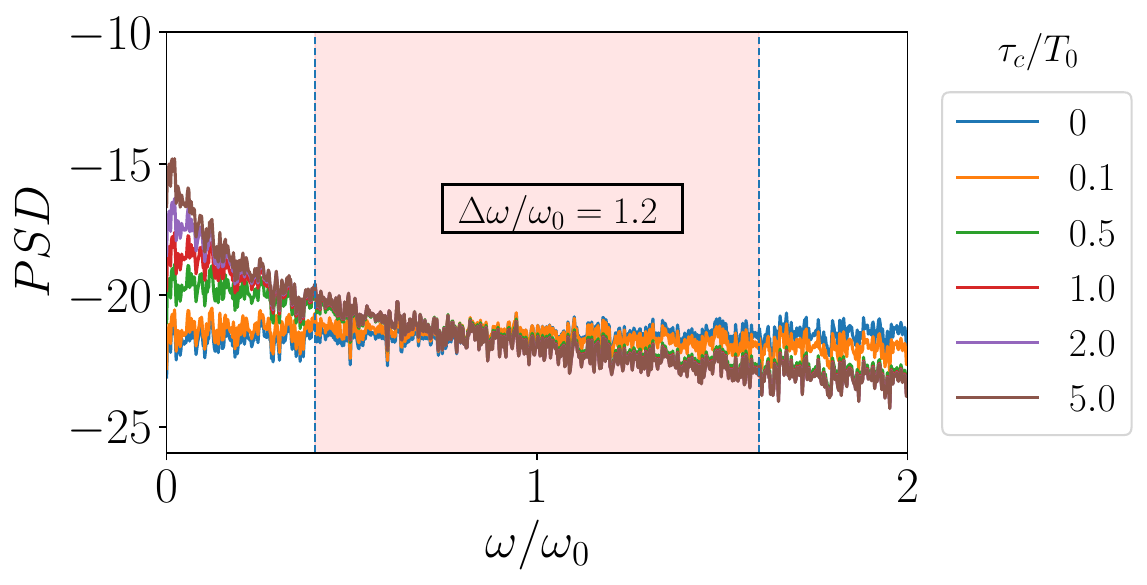}
    \caption{\justifying Power spectral density (PSD) of white noise $(\tau_c/T_0 = 0)$ and OU noises with varying correlation times at $\sigma = 0.05$. The OU noise is generated such that the powers provided by $\xi(t)$ and $\epsilon_\sigma(t)$ within the band $\Delta \omega / \omega_0 = 1.2$ are equal. $\tau_c$ is normalized by the duct acoustic time period $T_0 = 2 \pi / \omega_0$ at $K = 0.8$. The spectra show the low-pass nature of the colored noise, where higher frequencies are increasingly suppressed as the correlation time increases.}
    \label{noise}
\end{figure}

Although the noise amplitude can be varied independently, the bandwidth over which the noise power is distributed significantly influences the system response. If the chosen bandwidth is too narrow (for example, $\Delta\omega/\omega_0 =(\omega_2-\omega_1)/\omega_0=0.1$), information in the time series may be lost, while a large bandwidth (for example, $\Delta\omega/\omega_0 =5$) may include multiple neighboring spectral peaks when multiple eigenmodes are simultaneously excited, both leading to inaccurate results \cite{bonciolini2017output, vishnoi2024effect}. Thus, the generation of stochastic forcing requires careful selection of the spectral bandwidth around the system's natural frequency. So we follow the same bandwidth selection approach employed in earlier studies \cite{vishnoi2024effect,vishnoi2024reliability} and choose $\Delta\omega/\omega_0 = 1.2$, which effectively isolates the fundamental acoustic mode while minimizing interference from higher-order modes. The corresponding power spectra of the independently generated white and OU noises for tube A are shown in Fig.~\ref{noise}. The noise correlation time $\tau_c$ is normalized by the acoustic time period $T_0 = 2\pi/\omega_0$ evaluated at $K=0.8$. In this study, the normalized correlation time is taken as $\tau_c/T_0 = 0,0.1,0.5,1,2$ and $5$, extending up to twice the duct acoustic time period to capture physically relevant correlations, with an additional higher value included to examine a strongly correlated extreme case.

\section{Results and Discussion}\label{section_3}

In this section, we investigate the influence of stochastic forcing on the dynamics of two coupled thermoacoustic oscillators interacting through time-delay and dissipative coupling (Eqs. \ref{eq1} and \ref{eq2}). We first analyze the effect of colored noise on time-delay coupled oscillators. Subsequently, we consider purely dissipative coupling by switching off the delay coupling. For both coupling schemes, we examine the variation of $P_{rms}'$ with the corresponding coupling parameters. To comprehensively analyze how noise affects the AD region of coupled Rijke oscillators, we construct heatmaps in the $(\tau, K_\tau)$ parameter space for time-delay coupling and in the $(\omega_B/\omega_A, K_d)$ parameter space for dissipative coupling, under different noise intensities and correlation times. Here, we define AD as the regime in which the oscillation amplitude decreases to approximately $20\%$ of its initial uncoupled limit-cycle amplitude, following Thomas et al.~\cite{thomas2018effect}. Finally, we examine the coherence factor $\beta$ as a function of noise amplitude $\sigma$, heater power $K$, and correlation time $\tau_c$ for both coupling configurations. Unless otherwise specified, both oscillators are operated in the limit-cycle regime with $K_A = K_B = K = 0.8$.

Throughout this work, the coupled systems use the same system parameters as those used for the single Rijke-tube oscillator. The detailed numerical algorithm used for solving the coupled system is provided in Appendix \ref{algorithm}. The one- and two-parameter response plots are constructed from finite-time numerical simulations. The reported $P_{rms}'$ corresponds to the ensemble mean obtained from independent noises. The one-parameter response plots (Figs. \ref{time_delay_coupling_strength}, \ref{time_delay_tau}, and \ref{dissi_omega}) are performed for $\sigma = 0.05,\ 0.1,\ 0.2$ and are obtained from ensemble averages over 50 realizations. The two-parameter response plots (Figs. \ref{time_delay_full} and \ref{dis_both}) include an additional value $\sigma =0.4$ to capture the further evolution of the AD regions, and are obtained from ensemble averages over 10 realizations, because of the substantially higher computational cost. The chosen $\sigma$ range was intended to span weak, intermediate, and strong stochastic forcing relative to the characteristic thermoacoustic oscillation amplitude. The coherence factor analysis (Figs. \ref{coherence_Kt} and \ref{coherence_Kd}) is also performed using ensemble averages over 50 realizations. The associated statistical uncertainty is quantified using the standard error of the ensemble mean, $SE =s/\sqrt{N_{ens}}$, where $s$ is the sample standard deviation and $N_{ens}=50$. The uncertainties are shown for representative cases in Figs. \ref{time_delay_coupling_strength}, \ref{time_delay_tau}, \ref{coherence_Kt}, \ref{dissi_omega} and \ref{coherence_Kd}.

\subsection{Two coupled Rijke tubes under time delay coupling}\label{section_3A}
\subsubsection{Stochastic amplitude response}\label{bifur_time}

\begin{figure}
    \centering
    \includegraphics[width=\linewidth]{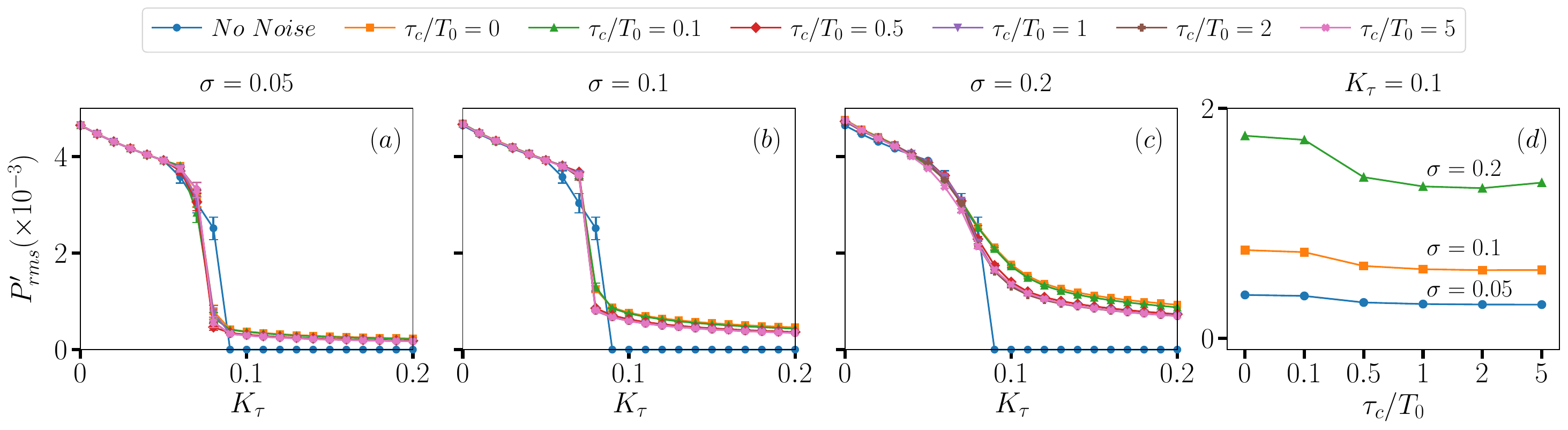}
    \caption{\justifying Variation in $P_{rms}'$ as a function of time-delay coupling strength $K_\tau$ in the deterministic case and under noise with different correlation times ($\tau_c / T_0$) for three noise amplitudes; (a) $\sigma = 0.05$, (b) $\sigma = 0.1$, (c) $\sigma = 0.2$}. (d) The $P_{rms}'$ for all noise color and amplitudes at $K_{\tau} = 0.1$.  For all plots, we take $\tau = 0.5$, and $K_A = K_B = 0.8$. Error bars represent the standard error of the ensemble mean of $P_{rms}'$.
    \label{time_delay_coupling_strength}
\end{figure}

\indent We first examine the influence of noise on AD under time-delay coupling alone ($K_\tau \neq 0$ and $K_d = 0$), for identical coupled oscillators $A$ and $B$ ($\omega^A  = \omega^B$). Figure~\ref{time_delay_coupling_strength} shows $P_{rms}'$ of oscillator $A$ as a function of the $K_\tau$ at a fixed delay $\tau = 0.5$, for three noise amplitudes $(\sigma = 0.05,\ 0.1,\ 0.2)$. In the deterministic case, the system undergoes an abrupt transition from LCO to AD at $K_\tau = 0.09$. At $\sigma=0.05$, the transition happens at $K_\tau = 0.08$ and retains much of its deterministic sharpness (Fig.~\ref{time_delay_coupling_strength}a), indicating that noise causes an earlier onset of amplitude suppression. At $\sigma = 0.1$ (Fig.~\ref{time_delay_coupling_strength}b), small-amplitude oscillations persist beyond the threshold point. At $\sigma = 0.2$, the transition becomes smooth, with larger-amplitude residual oscillations sustained well within the parameter range corresponding to AD in the noise-free case (Fig.~\ref{time_delay_coupling_strength}c). However, in all three noise amplitudes considered, the white and colored noise have negligible differences in the threshold, with only minute differences in the amplitude of the induced oscillations for the $\sigma = 0.2$ case. This difference is shown clearly in Fig.~\ref{time_delay_coupling_strength}d, where the $P_{rms}'$ is plotted for all noise color and amplitudes at fixed $K_\tau = 0.1$. The error bars show the statistical robustness of the observed results, with negligible deviation from the ensemble mean of $P_{rms}'$.

\begin{figure}
    \centering
    \includegraphics[width=\linewidth]{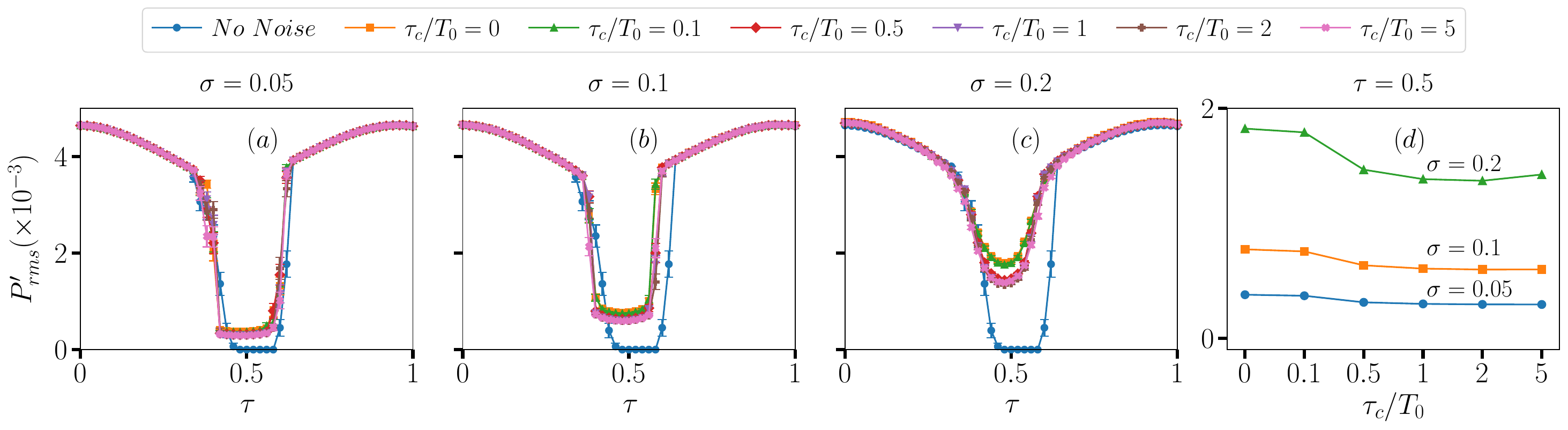}
    \caption{\justifying Variation in the $P_{rms}'$ with delay time $\tau$ under time-delay coupling in the deterministic case and under stochastic forcing with different correlation times for three noise amplitudes (a) $\sigma = 0.05$, (b) $\sigma = 0.1$, (c) $\sigma = 0.2$}, keeping fixed  $K_{\tau} = 0.1$. (d) The $P_{rms}'$ for all noise color and amplitudes at $\tau = 0.1$. The shown error bars represent the standard error of the ensemble mean of $P_{rms}'$.
    \label{time_delay_tau}
\end{figure}

We next examine the dependence of $P_{rms}'$ on the delay time $\tau$ for a fixed 
$K_\tau=0.1$ (Fig.~\ref{time_delay_tau}). In the deterministic case, the system undergoes similar abrupt transition from LCO to AD at $\tau = 0.48$ and the AD state persists until $\tau= 0.60$, after which the system returns to LCO, with the transitions at both boundaries of this AD window remaining relatively sharp. At $\sigma=0.05$, the transition remains largely as sharp as in the deterministic case (Fig.~\ref{time_delay_tau}a). At $\sigma = 0.1$, the onset of AD shifts marginally to $\tau = 0.44$, and small residual oscillations persist within the nominal AD window, indicating that complete amplitude suppression is no longer achieved (Fig.~\ref{time_delay_tau}b). At $\sigma = 0.2$ (Fig.~\ref{time_delay_tau}c), both transitions becomes quite smooth, and the residual oscillations within the suppression window attain larger amplitudes relative to the $\sigma=0.1$ case. Notably, LCO amplitude outside the suppression window remains largely unaffected across all noise intensities, suggesting that noise primarily disrupts the suppressed regime and erodes the sharpness of the transition boundaries rather than modifying the fully developed limit cycles. In Fig~\ref{time_delay_tau}d, the $P_{rms}'$ is plotted for all noise color and amplitudes at fixed $\tau = 0.1$, which shows the differences in the induced oscillation amplitude. The error bars remain very small across all noise intensities, with minute deviations from the ensemble mean occurring only in the vicinity of the transition regions.

\begin{figure}
    \centering
    \includegraphics[width=\linewidth]{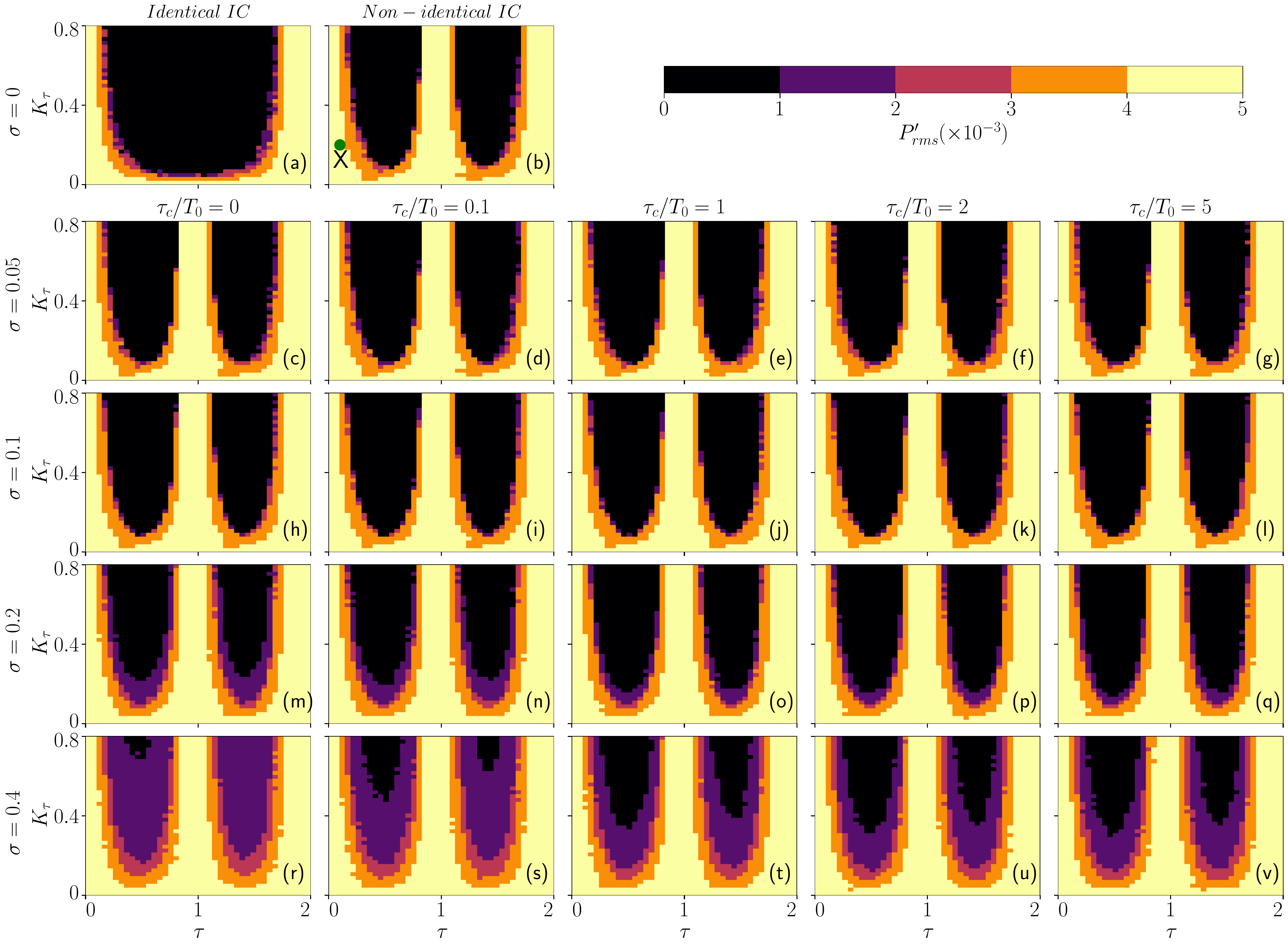}
    \caption{\justifying Contour maps of $P_{rms}'$ in $(\tau,K_\tau)$ plane under time-delay coupling. The first row corresponds to the deterministic case $(\sigma =0)$ when the two oscillators have (a) identical initial conditions, (b) different initial conditions, plots (c)-(v) show the response under stochastic forcing with different correlation times for four noise amplitudes: (c)-(g) $\sigma = 0.05$, (h)-(l) $\sigma = 0.1$, (m)-(q) $\sigma = 0.2$, (r)-(v) $\sigma = 0.4$. For the stochastic case, the five columns correspond to correlation times $\tau_c = 0$, $0.1$, $1$, $2$, and $5$. In all the plots, both oscillators are operated in the LCO regime of the control parameter, $K = 0.8$.}
    \label{time_delay_full}
\end{figure}

Further, we investigate the system response in the $(\tau, K_\tau)$ parameter plane without the addition of noise. The parameters $K_\tau$ and $\tau$ are varied in the ranges $0$ to $0.8$ and $0$ to $2$, respectively. The selected $\tau$ spans approximately one acoustic time period, while the $K_\tau$ range extends from the uncoupled limit to sufficiently strong coupling, so that the chosen domain adequately captures the transitions and the effects of coupling. The response to variations in the time-delay coupling parameters ($\tau, K_\tau$) depends on the symmetry of the initial conditions. We first consider the more general case in which the initial conditions are selected randomly for the two oscillators. The AD region exhibits two smaller, disconnected U-shaped regions centered approximately around $\tau = 1/2$ and $3/2$, as shown in Fig.~\ref{time_delay_full}b. The combinations of $(\tau, K_\tau)$ that lie outside these regions correspond to LCO. The boundary of these regions represent the minimum coupling strength required, at each delay time, to achieve AD. In contrast, when the oscillators have identical initial conditions, oscillator symmetry is preserved, and the system evolves towards the AD state within a single U-shaped region centered around $\tau = 1$, as shown in  Fig.~\ref{time_delay_full}a. A similar AD region has also been reported in \cite{thomas2018effect}. This behavior is not restricted to a particular initial state, provided that the initial conditions of both oscillators are identical (refer Appendix~\ref{appendix_phase}). It is worth noting that this exact symmetry-preserving condition is unlikely to be realized in practical experiments, where even small differences in the initial states or stochastic forcing of the oscillators can break the symmetry and lead to LCO. To support our observed numerical results, we then establish the deterministic stability boundaries of the time-delay coupled thermoacoustic system through linear stability analysis. The AD state corresponds to the stabilization of the zero-amplitude equilibrium, and its local stability is determined from the characteristic roots obtained by linearizing the coupled governing equations about this equilibrium. We verified the analytical and numerical Hopf points for the selected values of $K_\tau = 0.2$ and $\tau = 0.1$, and found good agreement with the numerical observations, provided in Appendix~\ref{stability_analysis_coupled}. We also examine analytically the initial-condition-dependent response observed in Figs.~\ref{time_delay_full}(a,b). At $K_\tau=0.3$ and $\tau=1$, corresponding to the AD response in Fig.~\ref{time_delay_full}(a) and LCO in Fig.~\ref{time_delay_full}(b), linear stability analysis of the quiescent-state yields  $\operatorname{Re}(\lambda_{\max})>0$. Thus, the AD state is unstable to small perturbations that breaks the symmetry of the two oscillators. These results confirm that the analytical stability analysis is consistent with the numerical results (refer to Appendix~\ref{stability_analysis_coupled} for the complete stability analysis).

When independent noise is introduced (Fig.~\ref{time_delay_full}c-v), the structure of the AD region closely resembles that of the Fig.~\ref{time_delay_full}(b) deterministic case, with separated U-shaped regions. At $\sigma = 0.05$, the overall topology and extent of the two AD regions remain largely unchanged for all noises (Fig.~\ref{time_delay_full}c-g). Even at $\sigma = 0.1$, the AD regions remain essentially unchanged (Fig.~\ref{time_delay_full}h-l). At $\sigma = 0.2$ (Fig.~\ref{time_delay_full}m-q), the two U-shaped AD regions persist, with a noticeable drop in their size and boundaries.
At $\sigma = 0.4$ (Fig.~\ref{time_delay_full}r-v), the influence of noise becomes more pronounced, which leads to a large reduction of the AD regions, requiring higher coupling strengths to achieve the prescribed $20\%$ amplitude suppression. Smaller the noise correlation time, larger the reduction in the AD region; white noise results in the largest reduction.

\subsubsection{Coherence Factor}\label{section_3C}
As the system approaches the saddle-node bifurcation point, the least stable acoustic mode becomes increasingly susceptible to stochastic excitation, giving rise to coherence resonance (CR). Near the bifurcation point, noise induces relatively regular oscillations even before the onset of self-sustained oscillations. To quantify this noise-induced coherence, we use the coherence factor $\beta$ \cite{neiman1997coherence,ushakov2005coherence} as a spectral measure of regularity in the coupled thermoacoustic system. CR is identified by a non-monotonic dependence of $\beta$ on the noise amplitude, with a peak at an intermediate noise amplitude indicating optimal coherence. The coherence factor is defined as 
\begin{equation*}
    \beta = H_p\Bigg(\frac{f_p}{\Delta f}\Bigg)
\end{equation*}
where $H_p$ is the height of the dominant spectral peak, $f_p$ is the peak frequency, and $\Delta f$ is the half-power bandwidth obtained from a Lorentzian fit of the peak. The term $f_p/\Delta f$ represents the spectral quality factor, and larger values of $\beta$ indicate stronger noise-induced coherence.

\begin{figure}
    \centering
    
    \begin{subfigure}[t]{0.48\textwidth}
        \centering
        \includegraphics[width=\linewidth]{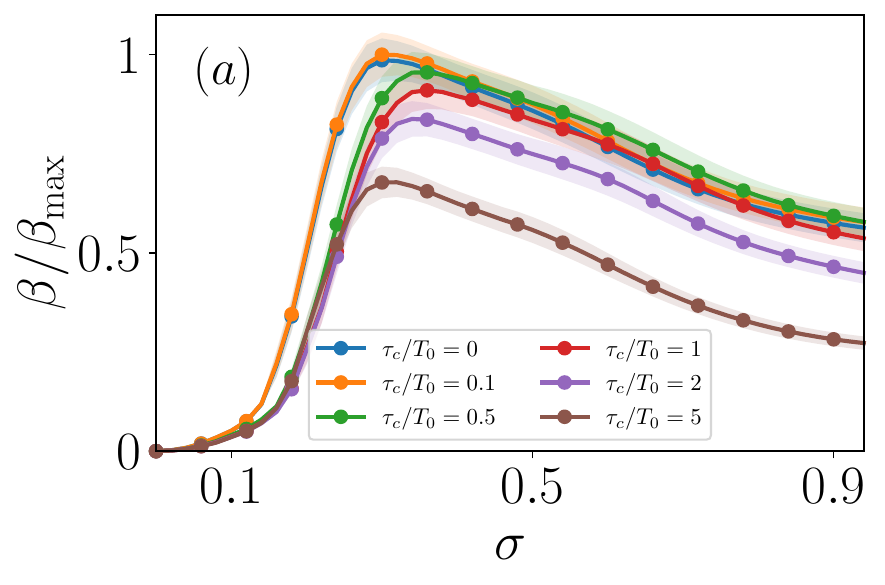}
        \label{coherence_intensity}
    \end{subfigure}
    \hfill
    \begin{subfigure}[t]{0.48\textwidth}
        \centering
        \includegraphics[width=\linewidth]{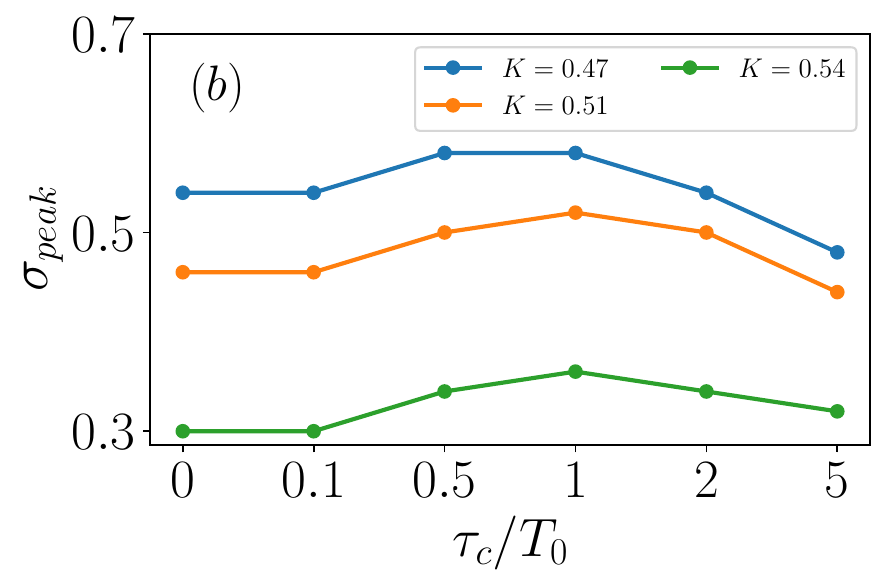}
        \label{coherence_max}
    \end{subfigure}

    \vspace{0.1em}
    \begin{subfigure}[t]{0.48\textwidth}
        \centering
        \includegraphics[width=\linewidth]{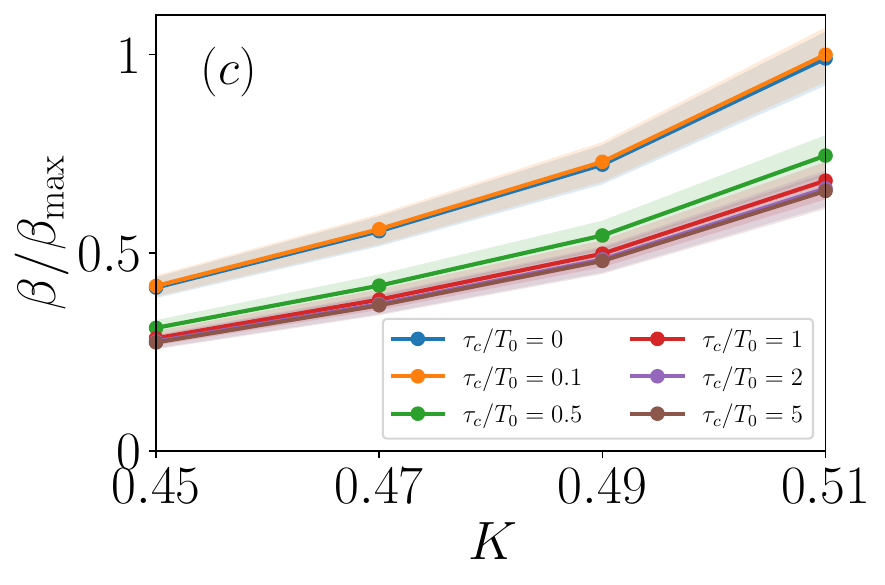}
        \label{coherence_heater_power}
    \end{subfigure}
    \hfill
    \begin{subfigure}[t]{0.48\textwidth}
        \centering
        \includegraphics[width=\linewidth]{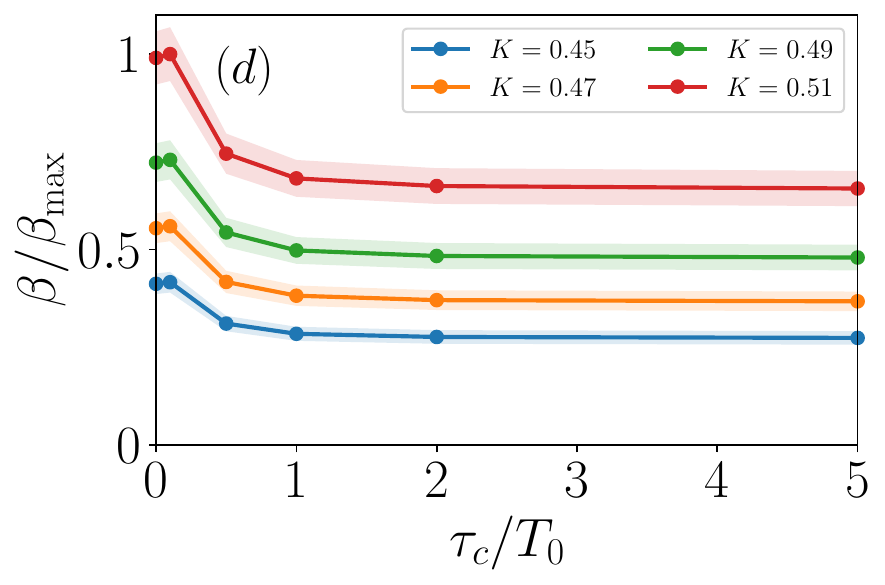}
        \label{coherence_correlation_time}
    \end{subfigure}
    
    \caption{\justifying 
     Normalized coherence factor ($\beta/\beta_{\max}$) under time-delay coupling ($K_\tau = 0.2$, $\tau = 0.1$). Dependence of $\beta/\beta_{\max}$ on; (a) noise amplitude $\sigma$ at $K = 0.54$, (b) variation of $\sigma_{peak}$ corresponding to the peak $\beta/\beta_{\max}$ as a function of $\tau_c/T_0$, (c) heater power $K$ near the saddle-node bifurcation at $\sigma = 0.05$, and (d) normalized noise correlation time $\tau_c/T_0$ for different values of $K$ at $\sigma = 0.05$. The shaded regions represent the standard error of the ensemble mean of $\beta/\beta_{max}$.}
    \label{coherence_Kt}

\end{figure}

To study the effect of noise on coherence resonance, we analyze the dependence of the normalized coherence factor $\beta/\beta_{max}$ on noise amplitude (Fig.~\ref{coherence_Kt}a), heater power (Fig.~\ref{coherence_Kt}c), and noise correlation time (Fig.~\ref{coherence_Kt}d) at $K_\tau = 0.2$ and $\tau = 0.1$ (green dot with X mark in Fig.~\ref{time_delay_full}b). Here, $\beta_{max}$ denotes the maximum value of $\beta$ of the observed results. The shaded regions in these figures represent the standard error of the mean obtained from ensemble. Note that $\beta \propto 1/(\Delta f)$, making the estimated coherence factor highly sensitive to the Lorentzian fit and the determination of the spectral bandwidth $\Delta f$, which contributes to the relative uncertainty reflected by the shaded regions. In Fig~\ref{coherence_Kt}a, the heater power is fixed at $K=0.54$, corresponding to a value just before to the shifted bifurcation point of the coupled system, as shown in Appendix~\ref{stability_analysis_coupled}. Here, $\beta/\beta_{max}$ increases with noise amplitude, reaches a maximum at an intermediate value, and then decreases, indicating optimal noise-induced coherence characteristic of coherence resonance. The peak coherence factor is highest for the correlation time much smaller than the acoustic time scale $(\tau_c/T_0=0.1)$. As the noise correlation time increases, the peak coherence factor becomes progressively lower and shifts to higher values of $\sigma$ till $\tau_c/T_0 = 1$, which then reduces for $\tau_c/T_0 > 1$.

Figure \ref{coherence_Kt}a shows that the noise amplitude for which peak coherence occurs ($\sigma_{\textrm{peak}}$) depends on noise color. This dependence is shown more clearly in Fig.~\ref{coherence_Kt}b, where $\sigma_{\textrm{peak}}$ is plotted against the noise color, for different values of the heater parameter, $K$, in the AD region. It is clear that for a given parameter value $K$, $\sigma_{\textrm{peak}}$ is a function of noise color and peaks at noise correlation time, $\tau_c = T_0$. Moreover, for a given noise color, $\sigma_{\textrm{peak}}$ increases as $K$ is reduced, i.e., as we move further into the AD region, away from the parameter value at which AD state transitions to LCO state.
Figure \ref{coherence_Kt}c shows that $\beta/\beta_{max}$ increases monotonically for all noises as the heater power is increased toward the saddle-node point, with noise amplitude fixed at $\sigma = 0.05$. This behavior reflects the reduced stability margin of the system near the bifurcation, where the least stable acoustic mode becomes increasingly susceptible to stochastic excitation. Figure~\ref{coherence_Kt}d illustrates the dependence of $\beta/\beta_{max}$ on $\tau_c$ for certain values of $K$, with the noise amplitude fixed at $\sigma = 0.05$. For small correlation times, $\beta/\beta_{max}$ decreases as $\tau_c$ increases. However, for $\tau_c$ beyond the acoustic time scale $T_0$, the coherence factor becomes constant. The observed trends remain qualitatively consistent across all values of $K_\tau$ and $\tau$ within the LCO regime of the two-parameter response map (Fig.~\ref{time_delay_full}), provided that the heater power $K$ remains below the saddle-node bifurcation point corresponding to the respective $(K_\tau,\tau)$ values.

\subsection{Two coupled Rijke tubes with dissipative coupling}\label{section_3B}
\subsubsection{Stochastic amplitude response} 

\indent In this subsection, we examine the influence of dissipative coupling ($K_{\tau}=0$ and $K_d \neq 0$) on amplitude suppression of the coupled thermoacoustic oscillators (Eq. \ref{eq2}), first in the deterministic limit and subsequently in the presence of noise. Previous studies suggest that dissipative coupling alone cannot induce AD in two identical oscillators $(\omega_B/\omega_A = 1)$. Therefore, to facilitate amplitude suppression through dissipative coupling, we introduce frequency detuning between the oscillators by considering $\omega_B/\omega_A \neq 1$.

\begin{figure}
    \centering
    \includegraphics[width=\linewidth]{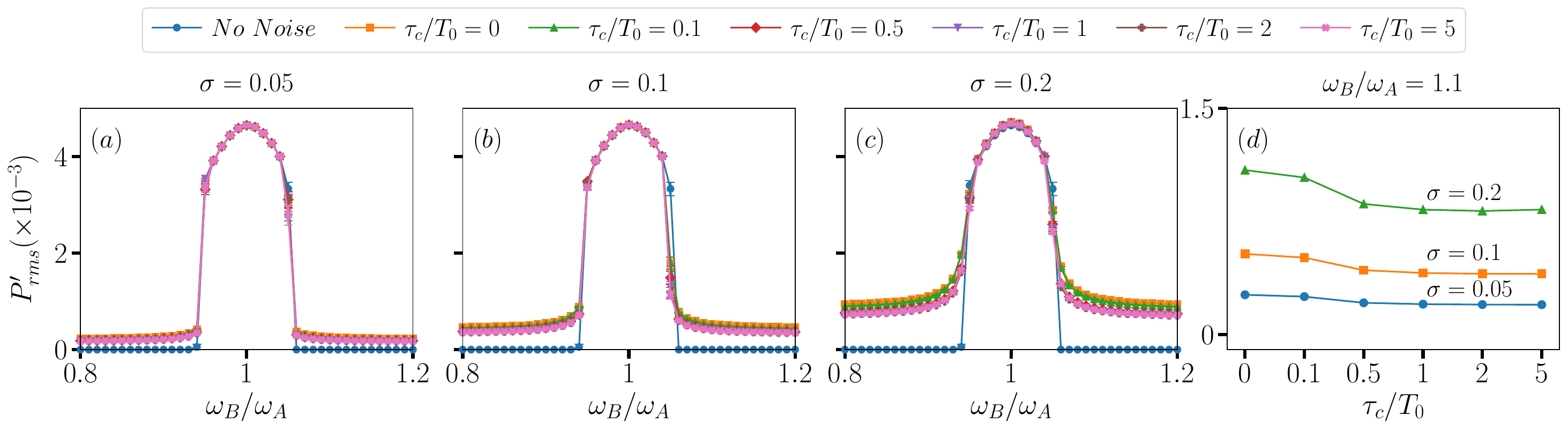}
    \caption{\justifying Variation in the $P_{rms}'$ with detuning $\omega_B/\omega_A$ under dissipative coupling ($K_d =0.2)$ in the deterministic case and under stochastic forcing with different correlation times for three noise amplitudes (a) $\sigma = 0.05$, (b) $\sigma = 0.1$, (c) $\sigma = 0.2$. (d) The $P_{rms}'$ for all noise color and amplitudes at $\omega_B/\omega_A = 1.1$. The sharp transitions on either side of $\omega_B/\omega _A =1$ becomes less abrupt as the noise amplitude increases. The shown error bars represent the standard error of the ensemble mean of $P_{rms}'$.}
    \label{dissi_omega}
\end{figure} 

We analyze the $P_{rms}'$ as a function of the detuning ratio $\omega_B/\omega_A$ (Fig.~\ref{dissi_omega}) at $K_d = 0.2$. In the deterministic case, AD is achieved only when the detuning ratio exceeds a critical threshold. As shown in Fig.~\ref{dissi_omega}, $P_{rms}'$ drops abruptly to zero beyond a critical detuning on either side of $\omega_B/\omega_A =1$, indicating sharp transitions from LCO to AD. This behavior highlights the stabilizing role of detuning, whereby exceeding a critical detuning drives the system to a stable AD state, suppressing oscillations. When noise is introduced, the qualitative AD structure is preserved, while the abrupt transition gets smoother for larger noise amplitudes. At $\sigma = 0.05$ and $ 0.1$ (Fig.~\ref{dissi_omega}a,b), the transition behavior remains identical to the deterministic case, with only small-amplitude oscillations in the AD region. However, at $\sigma = 0.2$ (Fig.~\ref{dissi_omega}c), the sharp transitions are significantly smoothed. In particular, complete cessation of oscillations is no longer attainable; instead, the suppressed regime exhibits residual oscillations of finite amplitude.  However, a small difference in the system amplitude is observed across the different noise types for the $\sigma = 2$ case, as shown in Fig.~\ref{dissi_omega}d, where $P_{rms}'$ is plotted at fixed $\omega_B/\omega_A=1.1$. The error bars show negligible deviations for all noise intensities and noise types except near transition points.For the dissipatively coupled system, we verified the analytical and numerical Hopf points for $K_d=0.4$ and $\omega_B/\omega_A=0.95$, finding good agreement between them (see Appendix~\ref{stability_analysis_coupled} for details).

\begin{figure}
    \centering
    \includegraphics[width=\linewidth]{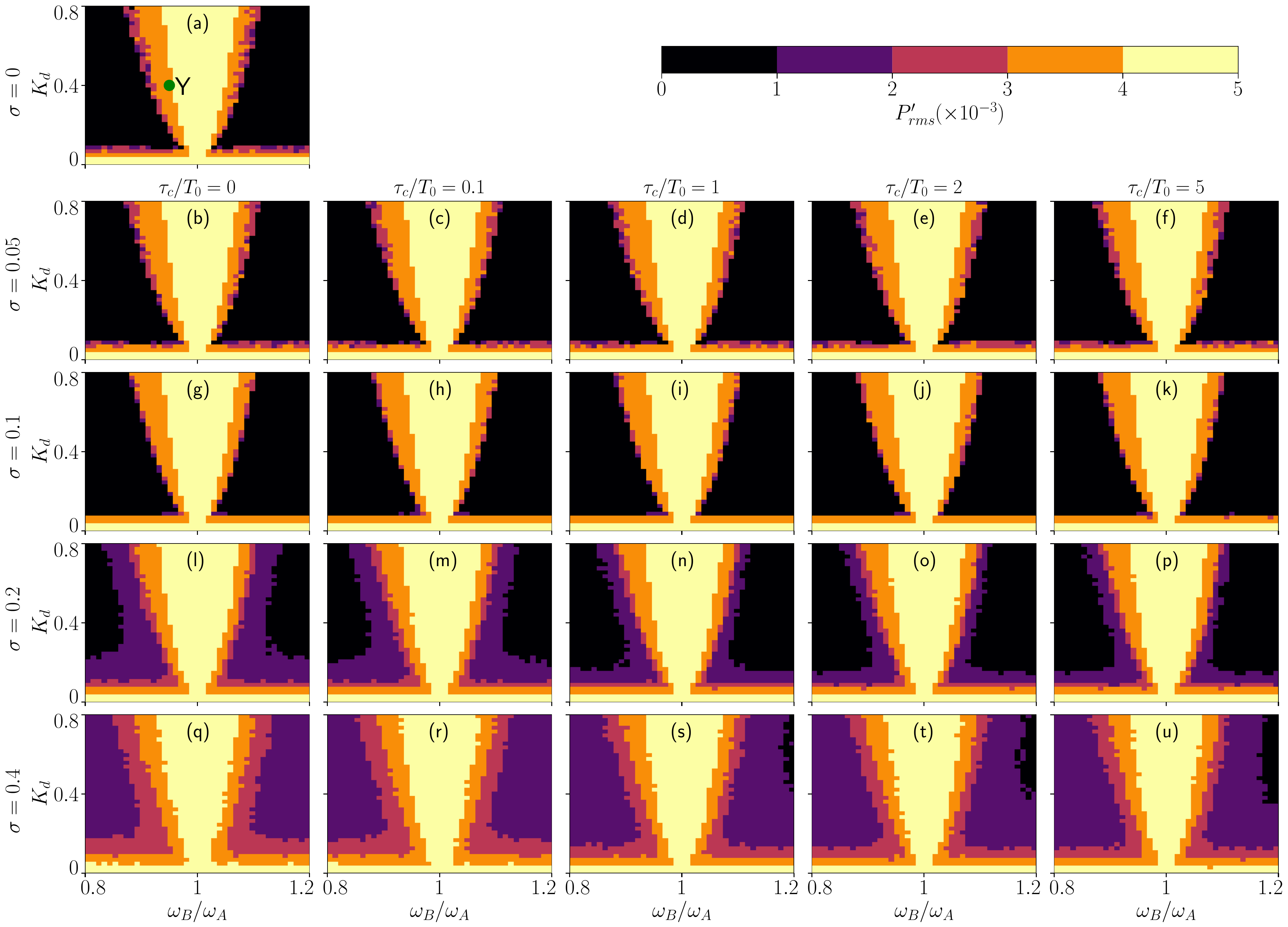}
    \caption{\justifying Contour maps of $P_{rms}'$ in $(\omega_B/\omega_A,K_d)$ plane under dissipative coupling. Plot (a) corresponds to the deterministic case $(\sigma =0)$, while plots (b)-(u) present the response under stochastic forcing with different correlation times for four noise intensities: (b)-(f) $\sigma = 0.05$, (g)-(k) $\sigma = 0.1$, (l)-(p) $\sigma = 0.2$, (q)-(u) $\sigma =0.4$. For the stochastic case, the five columns correspond to correlation times $\tau_c = 0$, $0.1$, $1$, $2$ and $5$. In all the cases, both oscillators are operated in the limit-cycle regime at $K = 0.8$.}
    \label{dis_both}
\end{figure} 

\begin{figure}
    \centering
    
    \begin{subfigure}[t]{0.48\textwidth}
        \centering
        \includegraphics[width=\linewidth]{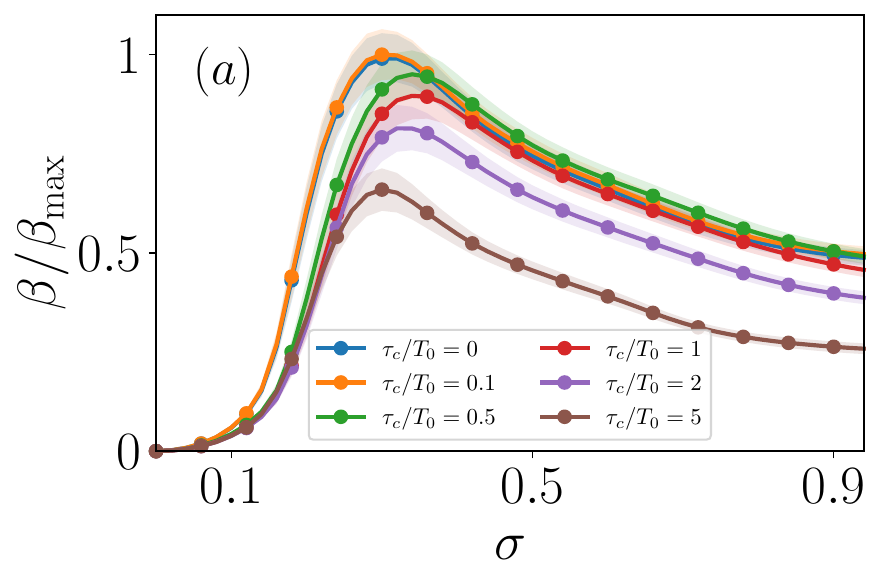}
        \label{coherence_intensity_a}
    \end{subfigure}
    \hfill
    \begin{subfigure}[t]{0.48\textwidth}
        \centering
        \includegraphics[width=\linewidth]{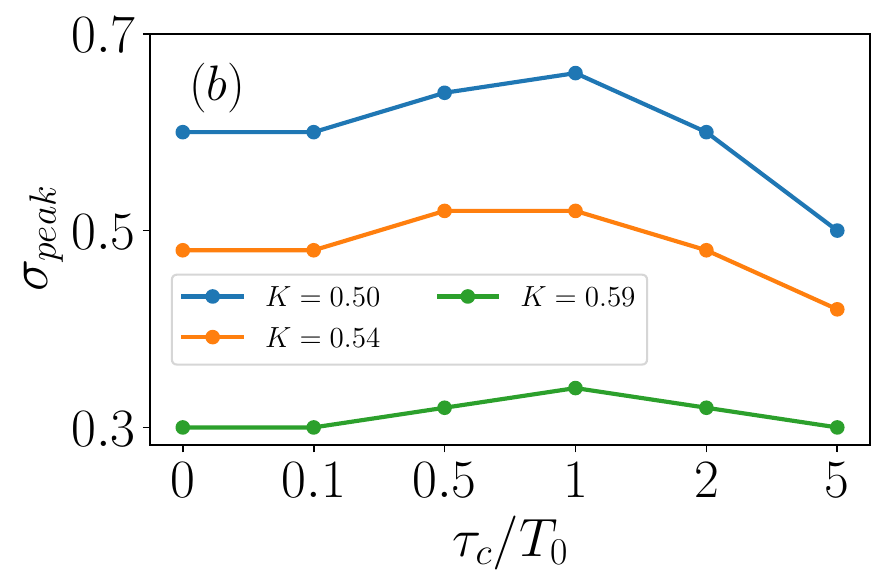}
        \label{coherence_maxA}
    \end{subfigure}
    
    \vspace{0.1em}
    \begin{subfigure}[t]{0.48\textwidth}
        \centering
        \includegraphics[width=\linewidth]{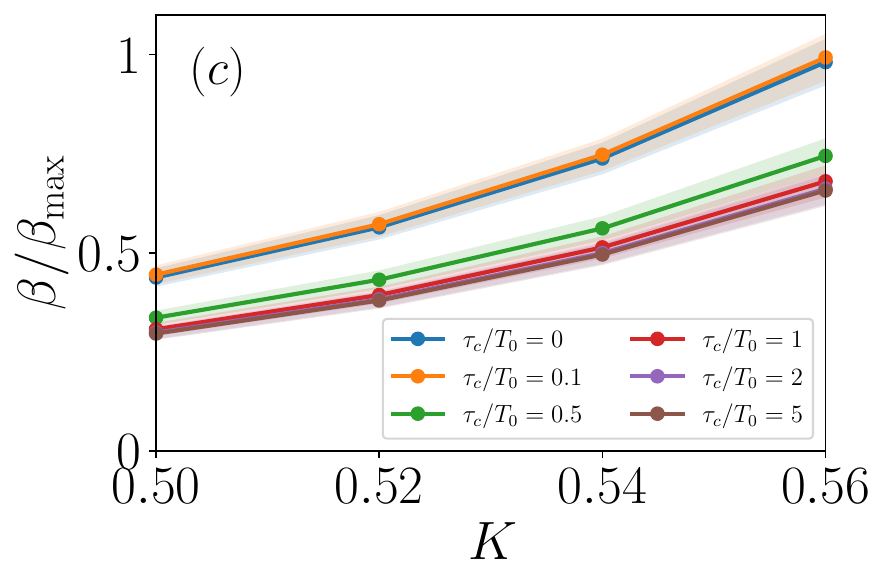}
        \label{coherence_heater_power_a}
    \end{subfigure}
    \hfill
    \begin{subfigure}[t]{0.48\textwidth}
        \centering
        \includegraphics[width=\linewidth]{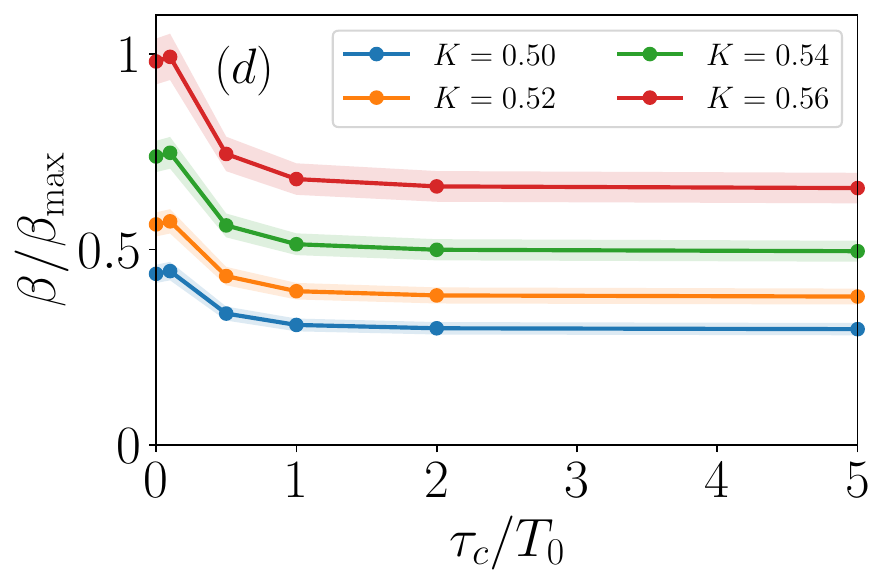}
        \label{coherence_correlation_time_a}
    \end{subfigure}

     \caption{\justifying Normalized coherence factor ($\beta/\beta_{\max}$) under dissipative coupling ($K_d = 0.4$,\ $\omega_B/\omega_A = 0.95$). Dependence of $\beta/\beta_{\max}$ on (a) noise amplitude $\sigma$ at $K = 0.59$, (b) variation of $\sigma_{peak}$ corresponding to the peak $\beta/\beta_{\max}$ as a function of $\tau_c/T_0$, (c) heater power $K$ as the system approaches the saddle-node bifurcation at $\sigma = 0.05$, and (d) noise correlation time $\tau_c$ for different values of $K$ at $\sigma = 0.05$. The shaded regions represent the standard error of the ensemble mean of $\beta/\beta_{max}$.}
    \label{coherence_Kd}
\end{figure} 

We further examine the response of the coupled system in the $(K_d, \omega_B/\omega_A)$ plane to observe the influence of noise $(K_\tau = 0)$ (Fig.~\ref{dis_both}). The first row corresponds to the deterministic case, while the remaining rows correspond to noise. In the deterministic case (Fig.~\ref{dis_both}a), AD regions appear on either side of $\omega_B/\omega_A = 1$, consistent with the requirement of sufficient detuning for dissipative coupling to suppress oscillations. For very small values of $K_d$, AD is not observed over the entire range of $\omega_B/\omega_A$. As $K_d$ increases beyond a critical value, the AD regime emerges at larger detuning values. The LCO region extends along the detuning axis but remains confined to a finite interval of $\omega_B/\omega_A$. At $\sigma = (0.05, 0.1)$ (Fig.~\ref{dis_both}b–k), the AD and LCO regions remain identical to those in the deterministic case. At $\sigma = 0.2$ (Fig.~\ref{dis_both}l–p), the AD region shrinks noticeably along its boundaries, with the smallest AD regions observed for white noise and colored noise with the shortest correlation time. At $\sigma = 0.4$ (Fig.~\ref{dis_both}q–u), the AD region disappears completely for the white noise and shortest correlation noise, while only small AD regions persist for longer correlation noise. Overall, the results under dissipative coupling exhibit behavior qualitatively similar to that observed for time-delay coupling.


\subsubsection{Coherence Factor}
We study the effect of noise on the coherence factor by fixing $K_d = 0.4$ and $\omega_B/\omega_A = 0.95$ (green dot with Y mark in Fig.~\ref{dis_both}a) under dissipative coupling. Figure \ref{coherence_Kd} presents the variation of the coherence factor $\beta/\beta_{max}$ with noise amplitude, heater power, and noise correlation time. In Fig.~\ref{coherence_Kd}a, the heater power is fixed at $K = 0.59$, chosen due to a similar shift of the Saddle-node point to a higher $K$ value, as in the time-delay case, which is shown in Appendix \ref{stability_analysis_coupled}. The qualitative behavior closely mirrors that observed for time-delay coupling (Fig.~\ref{coherence_Kt}). In particular, as shown in Fig.~\ref{coherence_Kd}a, $\beta/\beta_{max}$ reaches a maximum at an intermediate noise amplitude. The highest peak is observed for colored noise with shortest correlated time ($\tau_c/T_0=0.1$). In Fig.~\ref{coherence_Kd}b, we observe that the noise amplitude $\sigma_{peak}$ required to attain peak value of $\beta/\beta_{max}$ is highest for $\tau_c/T_0=1$. Figure \ref{coherence_Kd}c shows a monotonic increase in $\beta/\beta_{max}$ as the system approaches the saddle-node bifurcation. This trend is also consistent with the time-delay case. In Fig.~\ref{coherence_Kd}d, for a fixed noise amplitude $\sigma = 0.05$, $\beta/\beta_{max}$ decreases and becomes constant once the correlation time is comparable to the acoustic time scale. The same behavior is observed for other values of $K_d$ and $\omega_B/\omega_A$, confirming that the effect of noise amplitude and temporal correlation on spectral coherence is robust and largely independent of the specific coupling parameters.

\section{Conclusions}\label{section_4}

In this study, we investigate the effects of colored noise on the dynamical behavior and coherence resonance emerges in two coupled Rijke tube oscillators with time-delay and dissipative couplings. Stochastic forcing is modeled using an Ornstein-Uhlenbeck (OU) process. To enable a quantitative comparison between white and colored noise, we generate the noise such that the OU and white-noise forcings have equal power within a frequency band around the natural frequency of the system. We consider the noise amplitudes $\sigma=0.05,\ 0.1,\ 0.2$ corresponding approximately to $5\%$, $10\%$, and $20\%$ of the reference white noise amplitude, respectively. 
The normalized correlation time is varied over $\tau_c/T_0=0,\,0.1,\,0.5,\,1,\,2,$ and $5$, spanning the white-noise limit $\tau_c =0$, short-correlated noise ($\tau_c< T_0$), correlation times comparable to the characteristic acoustic period ($\tau_c\sim T_0$), and strongly correlated noise ($\tau_c>T_0$).

Our results show that, for both time-delay and dissipative coupling, the influence of noise correlation time on the pressure amplitude response depends strongly on the noise amplitude. At low noise amplitudes ($\sigma = 0.05,\ 0.1$), variations in the correlation time have a negligible influence on the $P'_{\mathrm{rms}}$ response. As the noise amplitude increases, the influence of temporal correlation becomes more pronounced, particularly near the transition between LCO and AD states. For $\sigma \geq 0.2$, noise smoothens the transition from LCO to AD and induces residual oscillations within the nominal AD regime. Among the correlation times considered, white noise and short-correlated colored noise ($\tau_c/T_0 = 0.1$) produce the largest residual oscillation amplitudes. Consequently, they lead to a greater reduction of the AD region within the investigated $(K_\tau,\tau)$ and $(K_d,\omega_B/\omega_A)$ parameter spaces for time-delay and dissipative coupling, respectively. Also, we analytically predicted Hopf boundaries of the deterministic system, which agree well with the numerical results, validating the stability analysis of the AD state.

The coherence factor analysis reveals three general features common to both coupling mechanisms. First, as the noise amplitude is varied, the coherence factor exhibits a maximum at an intermediate noise amplitude, indicating coherence resonance. The highest peak coherence factor is obtained for colored noise with a correlation time much shorter than the acoustic time scale ($\tau_c/T_0=0.1$). The noise amplitude corresponding to the peak coherence factor, $\sigma_{\mathrm{peak}}$, depends on the noise correlation time and reaches its maximum when the correlation time is comparable to the acoustic time scale ($\tau_c/T_0=1$). Second, for a fixed noise amplitude, the coherence factor increases monotonically with heater power below the instability threshold. Third, for a fixed noise amplitude, the coherence factor decreases with increasing noise correlation time up to approximately the acoustic time scale ($\tau_c/T_0=1$), beyond which it remains nearly constant.

Overall, white noise and the shortest-correlated OU noise ($\tau_c/T_0 =0.1$) exert the strongest influence on both the amplitude response and coherence resonance, whereas longer-correlated noise has a progressively weaker effect. The nature of the coupling does not qualitatively alter the noise-induced responses, since both time-delay and dissipative couplings exhibit similar trends in $P_{rms}'$ response, the coherence factor, and their dependence on noise correlation time.
Our results also demonstrate that the noise-induced trends remain qualitatively similar across the two model configurations considered here, namely, time-delay-coupled identical oscillators and dissipatively coupled non-identical oscillators with frequency detuning. Future studies could explore other sources of non-identicality, including mismatches in heater power, damping, and geometric properties, as well as larger networks of coupled thermoacoustic oscillators. Analytical analysis to characterize the observed noise-induced dynamics also remains an important direction for future work.


\bmhead{Acknowledgements}
CM acknowledges support from the Anusandhan National Research Foundation (ANRF), India (Grant Number EEQ/2023/001080).

\section*{Declarations}

\bmhead{Conflict of interest} The authors declare that they have no conflict of interest.
\bmhead{Data availability} The codes used to perform the numerical simulations and generate the data presented in this study are available from the first and corresponding author upon reasonable request after publication.
\bmhead{Author contributions} R.R. led the data curation, formal analysis, investigation, validation, and writing of the original draft. Y.P. assisted with data curation, formal analysis, investigation, validation, and writing of the original draft. L.K., A.S., and C.M. contributed equally to conceptualization, methodology, supervision, validation, and manuscript review and editing. C.M. led funding acquisition, project administration, and resource provision. All authors reviewed and approved the final manuscript.

\begin{appendices}

\section{Numerical integration algorithm}\label{algorithm}
The coupled thermoacoustic equations are solved using a hybrid deterministic–stochastic numerical integration scheme. The deterministic modal equations governing the acoustic field are solved using the 4th order Runge-Kutta method, while the OU stochastic process is integrated using the Euler-Maruyama method. The initial conditions for the modal amplitudes are independently sampled from a uniform distribution, $X\sim\mathcal{U}(-10^{-4},10^{-4})$. Each simulation is performed over the time interval of $t=0$ to $5000$ with a integration step of $0.001$ to ensure that the system reaches its asymptotic state. The initial transient is discarded, and the RMS acoustic pressure is calculated over the final $20\%$ of the simulation time. The following numerical procedure is employed.

\begin{enumerate}

\item The modal amplitudes $\eta_j^m$ and modal velocities $\dot{\eta}_j^m$ are initialized for each oscillator $m\in\{A,B\}$.
\item Calibrate the reference white-noise intensity $\Gamma_0$.
\item Scale the reference white noise to the desired noise amplitude by $\sigma$ such that $\epsilon_\sigma = \sigma \epsilon_0$.
\item For a correlation time $\tau_c$, the normalization coefficient $D(\tau_c)$ is evaluated:
\begin{equation*}
D(\tau_c)
=
\frac{\tau_c(\omega_2-\omega_1)}
{\tan^{-1}(\omega_2\tau_c)-\tan^{-1}(\omega_1\tau_c)},
\end{equation*}
so that the OU and white-noise processes contain the same power within the frequency interval $[\omega_1,\omega_2]$ surrounding the natural frequency $\omega_0$ of the thermoacoustic oscillator.
\item Generate the OU noise increment using the Euler–Maruyama scheme for each oscillator independently,

\begin{equation*}
\xi^{\,k+1}
=
\xi^{\,k}
-
\frac{\Delta t}{\tau_c}\xi^{\,k}
+
\frac{\sqrt{D}}{\tau_c}
\sqrt{\Delta t}\,
\epsilon_\sigma^{\,k},
\end{equation*}

where $\epsilon_\sigma^{\,k} \sim \mathcal{N}(0,\sigma^2 \Gamma_0)$. 
\item For each time step $i = 1,2,...,N_t$:

\begin{enumerate}

\item Evaluate the delayed acoustic velocity at $t_i-t_f$ using interpolation for the nonlinear heat release rate. For time-delay coupling, also compute the delayed modal variables at $t_i-\tau$.
\item Compute the deterministic RK4 increments at $t_{i+1}=t_i+\Delta t$: 

\begin{equation}
\begin{aligned}
k_{1\eta} &= \dot{\eta}_i\,\Delta t, \nonumber\\
k_{1\dot{\eta}} &= f(\eta_i,\dot{\eta}_i,t_i)\,\Delta t, \nonumber\\
k_{2\eta} &= \left(\dot{\eta}_i+\frac{1}{2}k_{1\dot{\eta}}\right)\Delta t, \nonumber\\
k_{2\dot{\eta}} &= f\left(\eta_i+\frac{1}{2}k_{1\eta},
\dot{\eta}_i+\frac{1}{2}k_{1\dot{\eta}},
t_i+\frac{\Delta t}{2}\right)\Delta t, \nonumber\\
k_{3\eta} &= \left(\dot{\eta}_i+\frac{1}{2}k_{2\dot{\eta}}\right)\Delta t, \nonumber\\
k_{3\dot{\eta}} &= f\left(\eta_i+\frac{1}{2}k_{2\eta},
\dot{\eta}_i+\frac{1}{2}k_{2\dot{\eta}},
t_i+\frac{\Delta t}{2}\right)\Delta t, \nonumber\\
k_{4\eta} &= \left(\dot{\eta}_i+k_{3\dot{\eta}}\right)\Delta t, \nonumber\\
k_{4\dot{\eta}} &= f\left(\eta_i+k_{3\eta},
\dot{\eta}_i+k_{3\dot{\eta}},
t_i+\Delta t\right)\Delta t.
\end{aligned}
\label{rk4_stages}
\end{equation}

\item Update the OU noise increment $\xi_i(t)$

\item Update the state vector
\begin{align*}
\eta_{i+1} &= \eta_i + \frac{1}{6}\left(k_{1\eta} + 2k_{2\eta} + 2k_{3\eta} + k_{4\eta}\right), \nonumber\\
\dot{\eta}_{i+1} &= \dot{\eta}_i + \frac{1}{6}\left(k_{1\dot{\eta}} + 2k_{2\dot{\eta}} + 2k_{3\dot{\eta}} + k_{4\dot{\eta}}\right).
\end{align*}
\end{enumerate}
\item Repeat procedure 3 until the prescribed final integration time is reached.

\item After discarding the transient, the RMS acoustic pressure and coherence factor are evaluated from the steady-state signal. For stochastic simulations, the reported results correspond to the ensemble mean obtained from independent noise realizations. The associated uncertainty is quantified using the standard error of the ensemble mean.

\end{enumerate}

\section{Number of Acoustic Modes}\label{appendix_modes}
\begin{figure}[h]
    \centering
    \includegraphics[width=0.6\linewidth]{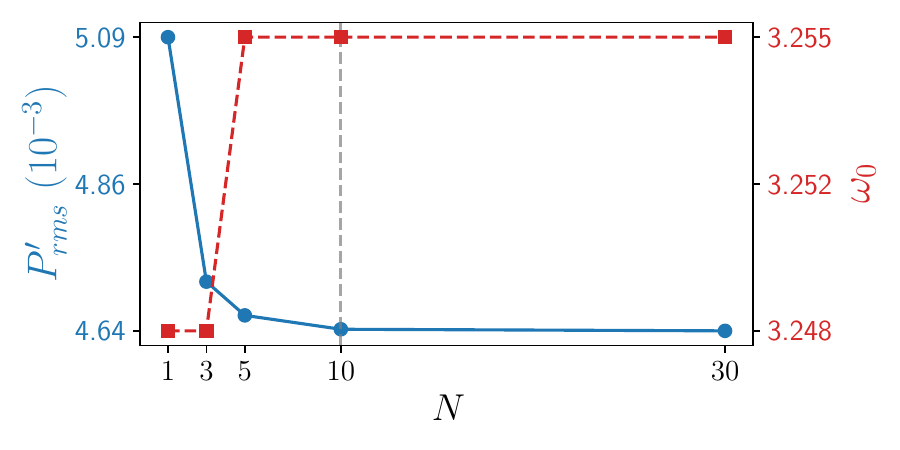}
    \caption{\justifying The convergence of $P'_{\mathrm{rms}}$ and $\omega_0$ with the increase in the number of Galerkin modes $N$ for a single Rijke tube at $K = 0.8$. The heater location is at $x_f = 0.25$ and the thermal time lag is fixed at $t_f = 0.2$.}
    \label{modes}
\end{figure}

To verify that the Galerkin truncation is adequate, we analyze how the system dynamics vary with the number of acoustic modes retained. Figure \ref{modes} shows the variation of the $P'_{\mathrm{rms}}$ and the corresponding non-dimensional natural frequency $\omega_0$ for different numbers of Galerkin modes $N$ at $K = 0.8$. As $N$ increases, both the oscillation amplitude and the dominant frequency approach steady values. In particular, increasing the number of modes beyond $N = 10$ produces no noticeable change in either quantity, indicating that retaining the first ten modes is sufficient to capture the essential thermoacoustic dynamics.

\section{Symmetry-Dependent Attractor}\label{appendix_phase}
\begin{figure}
    \centering
    \includegraphics[width=\linewidth]{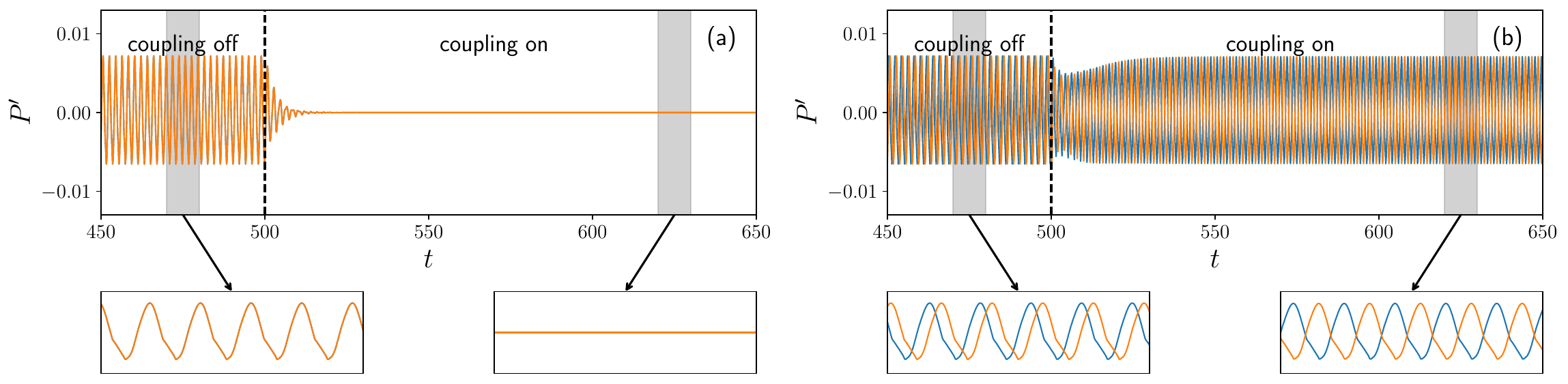}
    \caption{\justifying Time series of $P'$ illustrating noise induced AD and LCO for time-delay coupling at $K_{\tau} = 0.3$, $\tau = 1$, and $K = 0.8$. The coupling is introduced at $t = 500$ (vertical dashed lines). (a) Identical initial conditions preserve the symmetry between the two oscillator states and lead to AD. (b) Different initial conditions break the symmetry and lead to anti-phase LCO.}
    \label{time series no noise}
\end{figure}

In the vicinity of $\tau=1$ of the two-parameter deterministic response plots with time-delay coupling (Fig.~\ref{time_delay_full}a,b), two distinct dynamical states are observed:  AD and LCO. The observed response depends on the symmetry of the initial conditions of the two coupled oscillators. When the oscillators are initialized with identical initial conditions,  the symmetry between them is preserved. Despite the intrinsic tendency of each oscillator to sustain oscillations through thermoacoustic feedback, the coupled system exhibits a net loss of energy over an oscillation cycle, leading to AD in both oscillators (Fig.~\ref{time series no noise}a). In contrast, when the oscillators are initialized with slightly different initial conditions, with a difference of $10^{-6}$ (Fig.~\ref{time series no noise}b), their symmetry is broken and LCO is triggered. The two oscillators are locked-in to a constant phase difference of $\pi$.

\begin{figure}
    \centering
    \includegraphics[width=0.7\linewidth]{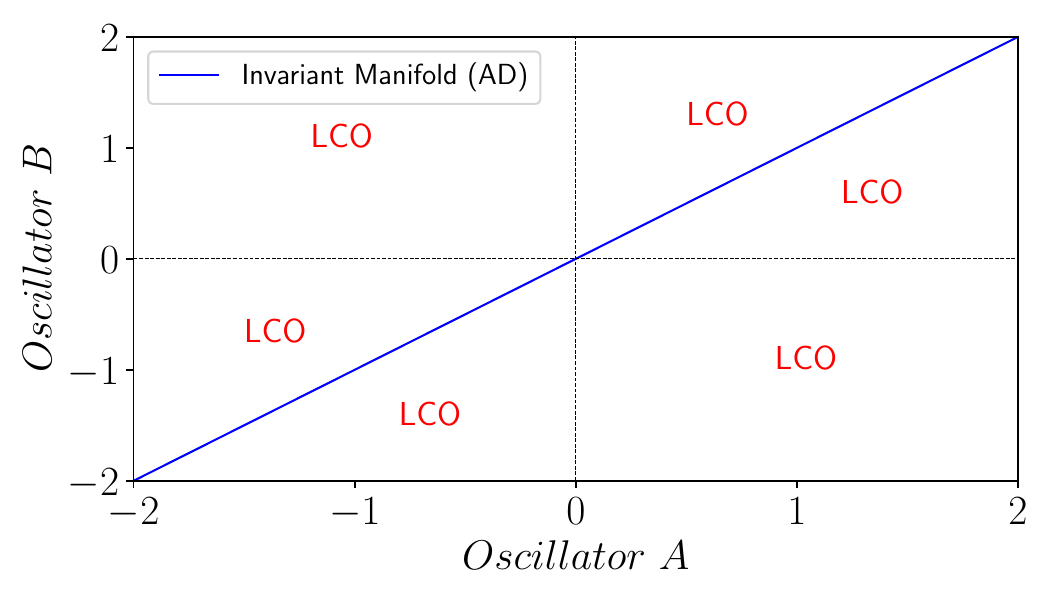}
    \caption{Initial-condition map illustrating the symmetry-dependent dynamical responses of two identical oscillators under time-delay coupling in the vicinity of \(\tau=1\). The blue diagonal represents the invariant manifold, defined by identical initial conditions of the two oscillators. Initial conditions on this manifold lead to AD, whereas initial conditions away from the manifold lead to anti-phase LCO.}
    \label{manifold}
\end{figure}

Such synchronization dynamics, characterized by an invariant manifold, has been previously reported for delay-coupled oscillators \cite{phase2010amplitude}. The invariant manifold is defined by the equality of states of the two identical oscillators at all times, such that trajectories initialized on this manifold preserve identical oscillator states. In the parameter regime considered for our system, such trajectories evolve towards AD. In contrast, initial conditions that do not belong to this manifold introduce a difference between the oscillator states and can lead to anti-phase LCO. The invariant manifold and the corresponding symmetry-breaking initial conditions are illustrated schematically in Fig.~\ref{manifold}. The figure shows the initial-condition map for the two oscillators near $\tau=1$ under time-delay coupling. Initial conditions on the diagonal (blue line) lead to AD, while small deviations from this manifold can result in anti-phase LCO.

\begin{figure}
    \centering
    \includegraphics[width=\linewidth]{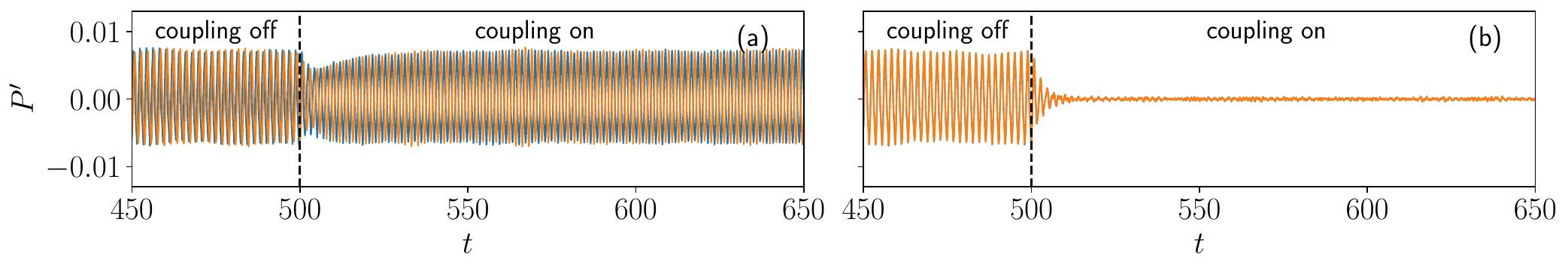}
    \caption{\justifying Time series of $P'$ for identical initial conditions of the time-delay coupled oscillators at $K_{\tau} =0.3$ and $\tau = 1$ under white noise of amplitude $\sigma =0.05$. (a) Different noise to both oscillators break the symmetry and leads to anti-phase LCO, (b) Same noise to both oscillators preserve the symmetry and leads to AD.}
    \label{time_series_noise_comparison}
\end{figure}

\begin{figure}
    \centering
    \includegraphics[width=\linewidth]{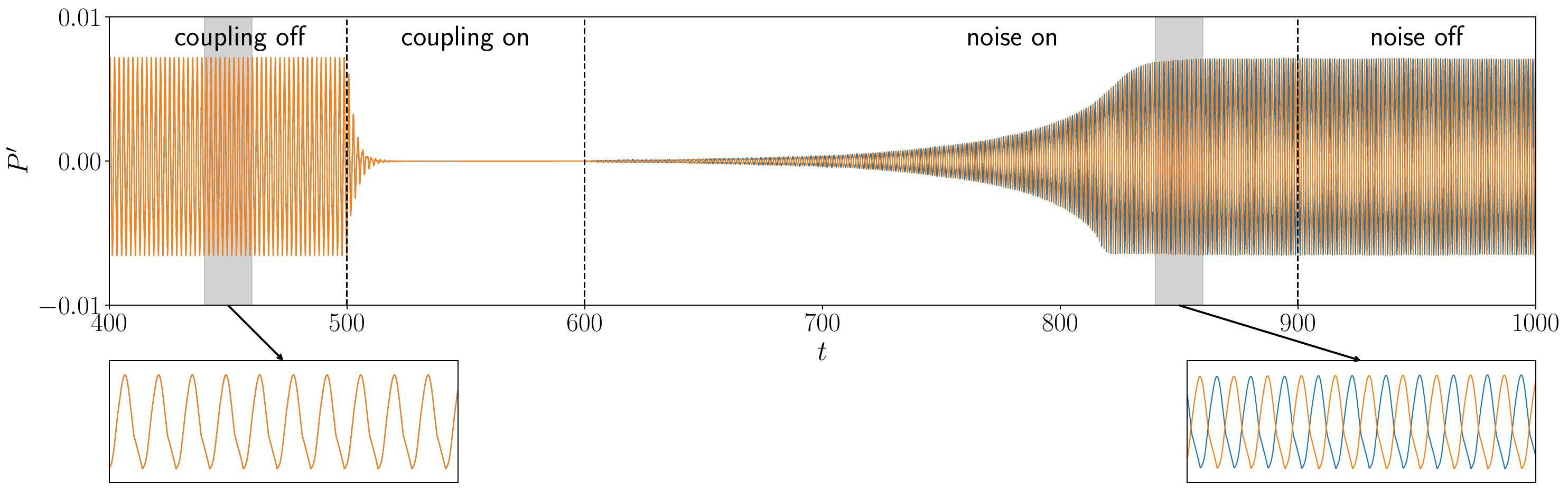}
    \caption{\justifying Time series of $P'$ for identical initial conditions of the coupled oscillators at $K_{\tau} =0.3$ and $\tau = 1$, showing AD induced by time-delay coupling at $t = 500$ in the absence of noise, followed by the re-emergence of LCO upon the introduction of white noise of amplitude $\sigma =0.05$ at $t= 600$. In the noise-induced LCO state, the two oscillators maintain a constant phase difference of $\pi$ (magnified section of the time series). The LCO persists even when the noise is removed from the system at $t= 900$. The heater parameter is set to $K = 0.8$.}
    \label{time_series}
\end{figure}

When two oscillators have identical initial conditions and independent stochastic forcing is applied to both oscillators, the system consistently evolves towards anti-phase LCO, as illustrated in Fig.~\ref{time_series_noise_comparison}a. Interestingly, a contrasting behavior is observed when oscillators are initialized with identical conditions but subjected to the same stochastic forcing Fig.~\ref{time_series_noise_comparison}b. In this case, the system invariably evolves towards the AD state, irrespective of the noise characteristics, including its intensity and correlation time, and the particular initial conditions selected. Even if the system has reach AD without noise, independent stochastic forcing can make the system to depart from the invariant manifold and evolve towards the anti-phase synchronized state, where sustained LCO emerge. Once the system reaches this attractor, the LCO persists even after the stochastic forcing is removed. Therefore, regardless of whether the coupled oscillators are initialized with identical or non-identical conditions, independent noises induce a transition to LCO. This behavior is illustrated in Fig.~\ref{time_series}, which shows the time series of pressure fluctuation at $K_{\tau} =0.3$ and $\tau = 1$, capturing the transition from LCO to AD due to coupling and from AD to sustained oscillations in the presence of noise. The LCO observed after noise removal indicates that the resulting oscillatory state is maintained by the deterministic dynamics. To quantify the observed response, we examined 50 independent stochastic realizations under white-noise forcing of amplitude $\sigma =0.05$, using the same initial conditions as in the deterministic case for both oscillators. All 50 realizations evolved towards anti-phase LCO, while none reached AD.

\section{Stability analysis of coupled Thermoacoustic system}\label{stability_analysis_coupled}

To determine the Hopf bifurcation associated with the transition between the AD and LCO states and to assess the stability of the AD state, we perform a linear stability analysis \cite{meena2023emergent,dawn2026origins, acharyya2026comprehensive} of the coupled thermoacoustic system governed by Eqs.~\ref{eq1} and \ref{eq2}. In the analysis, the two thermoacoustic oscillators are allowed to be non-identical.
In particular, we assume $\omega_j^A \neq \omega_j^B$ in general, allowing for the frequency mismatch relevant to the occurrence of AD in dissipatively coupled non-identical oscillators. The AD state is characterized by the stable non-oscillatory equilibrium
\begin{equation}
\eta_j^A=\eta_j^B=\dot{\eta}_j^A=\dot{\eta}_j^B=0,
    \qquad j=1,2,\ldots,N.
    \label{eq:AD_equilibrium}
\end{equation}
To determine the stability of the equilibrium states of the deterministic system, the stochastic forcing term in Eq.~\ref {eq2} is set to zero, and the governing equations are linearized about the non-oscillatory equilibrium states obtained from Eq.~\ref{eq:AD_equilibrium}.

The nonlinearity in the thermoacoustic model originates from the nonlinear dependence of the unsteady heat-release rate on the acoustic velocity fluctuation at the heater location. Specifically, the heat-release response is described by the modified King's law given in Eq.~\eqref{king's_law}. The corresponding nonlinear function can be expressed as \cite{sujith_bifurcation_analysis, srikanth2022}
\begin{equation}
    F(U_f')=\sqrt{\frac{1}{3}+U_f'(t-t_f)}-\sqrt{\frac{1}{3}}.
    \label{eq:nonlinear_heat_release_function}
\end{equation}
For the local stability analysis about the AD state, $|U_f'|\ll 1/3$. Expanding Eq.~\eqref{eq:nonlinear_heat_release_function} about $U_f'=0$ gives
\begin{equation}
    F(U_f')=\frac{\sqrt{3}}{2}U_f'(t-t_f)+\mathcal{O}\!\left((U_f')^2\right).
    \label{kings_approximation}
\end{equation}
Retaining only the linear contribution yields the linearized heat-release feedback.

Using the Galerkin expansion in Eq.~\eqref{galerkin} together with Eq.~\eqref{kings_approximation}, 
the linearized modal equations for tube $A$ can be written as
\begin{align}
    \frac{d\eta_j^A}{dt} &= \dot{\eta}_j^A, \label{eq:linear_eta_A}\\[2mm]
    \frac{d\dot{\eta}_j^A}{dt}
    +2\zeta_j^A\omega_j^A\dot{\eta}_j^A+(\omega_j^A)^2\eta_j^A \nonumber
    &=-\frac{\sqrt{3}}{2}j\pi K^A\sin(j\pi x_f^A)
    \sum_{k=1}^{N}\eta_k^A(t-t_f)\cos(k\pi x_f^A) \nonumber\\
    &\quad+K_d\left(\dot{\eta}_j^B-\dot{\eta}_j^A\right)
    +K_\tau\left[\dot{\eta}_j^B(t-\tau)-\dot{\eta}_j^A(t)\right].
    \label{eq:linear_modal_A}
\end{align}
Similarly, we can obtain for tube $B$ by interchanging the superscript $A$ and $B$. Here, $k=1,\ldots,N$ denotes the Galerkin mode index over which the modal contributions are summed, while $j$ identifies the particular Galerkin mode whose governing equation is being considered.
For compactness, we define
\begin{equation}
    B_{jk}^{m}=\frac{\sqrt{3}}{2}j\pi K^{m}\sin(j\pi x_f^{m})\cos(k\pi x_f^{m}),
    \qquad m\in\{A,B\}.
    \label{eq:B_matrix_general}
\end{equation}
Thus,
\begin{align}
    \ddot{\eta}_j^A
    &=-2\zeta_j^A\omega_j^A\dot{\eta}_j^A-(\omega_j^A)^2\eta_j^A
    -\sum_{k=1}^{N}B_{jk}^A\eta_k^A(t-t_f) \nonumber\\
    &\quad+K_d\left(\dot{\eta}_j^B-\dot{\eta}_j^A\right)
    +K_\tau\left[\dot{\eta}_j^B(t-\tau)-\dot{\eta}_j^A(t)\right],
    \label{eq:linear_second_A}
\end{align}
and
\begin{align}
    \ddot{\eta}_j^B
    &=-2\zeta_j^B\omega_j^B\dot{\eta}_j^B-(\omega_j^B)^2\eta_j^B
    -\sum_{k=1}^{N}B_{jk}^B\eta_k^B(t-t_f) \nonumber\\
    &\quad+K_d\left(\dot{\eta}_j^A-\dot{\eta}_j^B\right)
    +K_\tau\left[\dot{\eta}_j^A(t-\tau)-\dot{\eta}_j^B(t)\right].
    \label{eq:linear_second_B}
\end{align}

Introducing the modal vectors
\begin{equation}
    \boldsymbol{\eta}^{m}=\begin{bmatrix}\eta_1^m&\eta_2^m&\cdots&\eta_N^m\end{bmatrix}^{T},
    \qquad m\in\{A,B\},
\end{equation}
we define the $4N$-dimensional state vector
\begin{equation}
    \mathbf{X}(t)=
    \begin{bmatrix}
        \boldsymbol{\eta}^{A}(t)\\
        \boldsymbol{\eta}^{B}(t)\\
        \dot{\boldsymbol{\eta}}^{A}(t)\\
        \dot{\boldsymbol{\eta}}^{B}(t)
    \end{bmatrix}.
    \label{eq:state_vector}
\end{equation}
The linearized dynamics can then be written as
\begin{equation}
    \dot{\mathbf{X}}(t)=\mathbf{A}_0\mathbf{X}(t)
    +\mathbf{A}_1\mathbf{X}(t-t_f)
    +\mathbf{A}_2\mathbf{X}(t-\tau).
    \label{eq:compact_linear_system}
\end{equation}

Here, $\mathbf{A}_0$ contains the instantaneous dynamics and dissipative coupling, $\mathbf{A}_1$ represents the delayed thermoacoustic heat-release feedback, and $\mathbf{A}_2$ represents the delayed inter-tube coupling. They are given by
\begin{equation}
\mathbf{A}_0=
\begin{bmatrix}
\mathbf{0}&\mathbf{0}&\mathbf{I}&\mathbf{0}\\
\mathbf{0}&\mathbf{0}&\mathbf{0}&\mathbf{I}\\
-\boldsymbol{\Omega}_A^2&\mathbf{0}&-\mathbf{D}_A-(K_d+K_\tau)\mathbf{I}&K_d\mathbf{I}\\
\mathbf{0}&-\boldsymbol{\Omega}_B^2&K_d\mathbf{I}&-\mathbf{D}_B-(K_d+K_\tau)\mathbf{I}
\end{bmatrix},
\label{eq:A0}
\end{equation}
\begin{equation}
\mathbf{A}_1=
\begin{bmatrix}
\mathbf{0}&\mathbf{0}&\mathbf{0}&\mathbf{0}\\
\mathbf{0}&\mathbf{0}&\mathbf{0}&\mathbf{0}\\
-\mathbf{B}^{A}&\mathbf{0}&\mathbf{0}&\mathbf{0}\\
\mathbf{0}&-\mathbf{B}^{B}&\mathbf{0}&\mathbf{0}
\end{bmatrix},
\label{eq:A1}
\end{equation}
and
\begin{equation}
\mathbf{A}_2=
\begin{bmatrix}
\mathbf{0}&\mathbf{0}&\mathbf{0}&\mathbf{0}\\
\mathbf{0}&\mathbf{0}&\mathbf{0}&\mathbf{0}\\
\mathbf{0}&\mathbf{0}&\mathbf{0}&K_\tau\mathbf{I}\\
\mathbf{0}&\mathbf{0}&K_\tau\mathbf{I}&\mathbf{0}
\end{bmatrix}.
\label{eq:A2}
\end{equation}
Here, $\mathbf{0}$ and $\mathbf{I}$ denote the $N\times N$ zero and identity matrices, respectively. The frequency and damping matrices are
\begin{equation}
    \boldsymbol{\Omega}_m^2=\operatorname{diag}\left[(\omega_1^m)^2,(\omega_2^m)^2,\ldots,(\omega_N^m)^2\right],
    \qquad m\in\{A,B\},
    \label{eq:Omega_matrix}
\end{equation}
and
\begin{equation}
    \mathbf{D}_m=\operatorname{diag}\left[2\zeta_1^m\omega_1^m,2\zeta_2^m\omega_2^m,\ldots,2\zeta_N^m\omega_N^m\right],
    \qquad m\in\{A,B\}.
    \label{eq:D_matrix}
\end{equation}
The matrix $\mathbf{B}^{m}$ is the $N\times N$ thermoacoustic feedback matrix whose elements are defined by Eq.~\eqref{eq:B_matrix_general}.

To determine the local linear stability of the quiescent state, we seek exponential solutions of the form
\begin{equation}
    \mathbf{X}(t)=\mathbf{v}e^{\lambda t},
    \qquad \mathbf{v}\neq\mathbf{0},
    \label{eq:exponential_ansatz}
\end{equation}
where $\lambda\in\mathbb{C}$ denotes a characteristic root of the linearized delay system and $\mathbf{v}\in\mathbb{C}^{4N}$ is the corresponding characteristic vector. The delayed states are
\begin{equation}
    \mathbf{X}(t-t_f)=e^{-\lambda t_f}\mathbf{v}e^{\lambda t},
\end{equation}
and
\begin{equation}
    \mathbf{X}(t-\tau)=e^{-\lambda\tau}\mathbf{v}e^{\lambda t}.
\end{equation}
Substitution into Eq.~\eqref{eq:compact_linear_system} gives
\begin{equation}
    \left[\lambda\mathbf{I}_{4N}-\mathbf{A}_0
    -\mathbf{A}_1e^{-\lambda t_f}
    -\mathbf{A}_2e^{-\lambda\tau}\right]\mathbf{v}=\mathbf{0}.
    \label{eq:characteristic_root_problem}
\end{equation}
A non-trivial solution exists when
\begin{equation}
    \det\left[\lambda\mathbf{I}_{4N}-\mathbf{A}_0
    -\mathbf{A}_1e^{-\lambda t_f}
    -\mathbf{A}_2e^{-\lambda\tau}\right]=0.
    \label{eq:characteristic_equation}
\end{equation}
Owing to the exponential dependence on $\lambda$, Eq.~\eqref{eq:characteristic_equation} is a transcendental characteristic equation and, in general, possesses an infinite set of characteristic roots.

\begin{figure}
    \centering
    \includegraphics[width=1.0\linewidth]{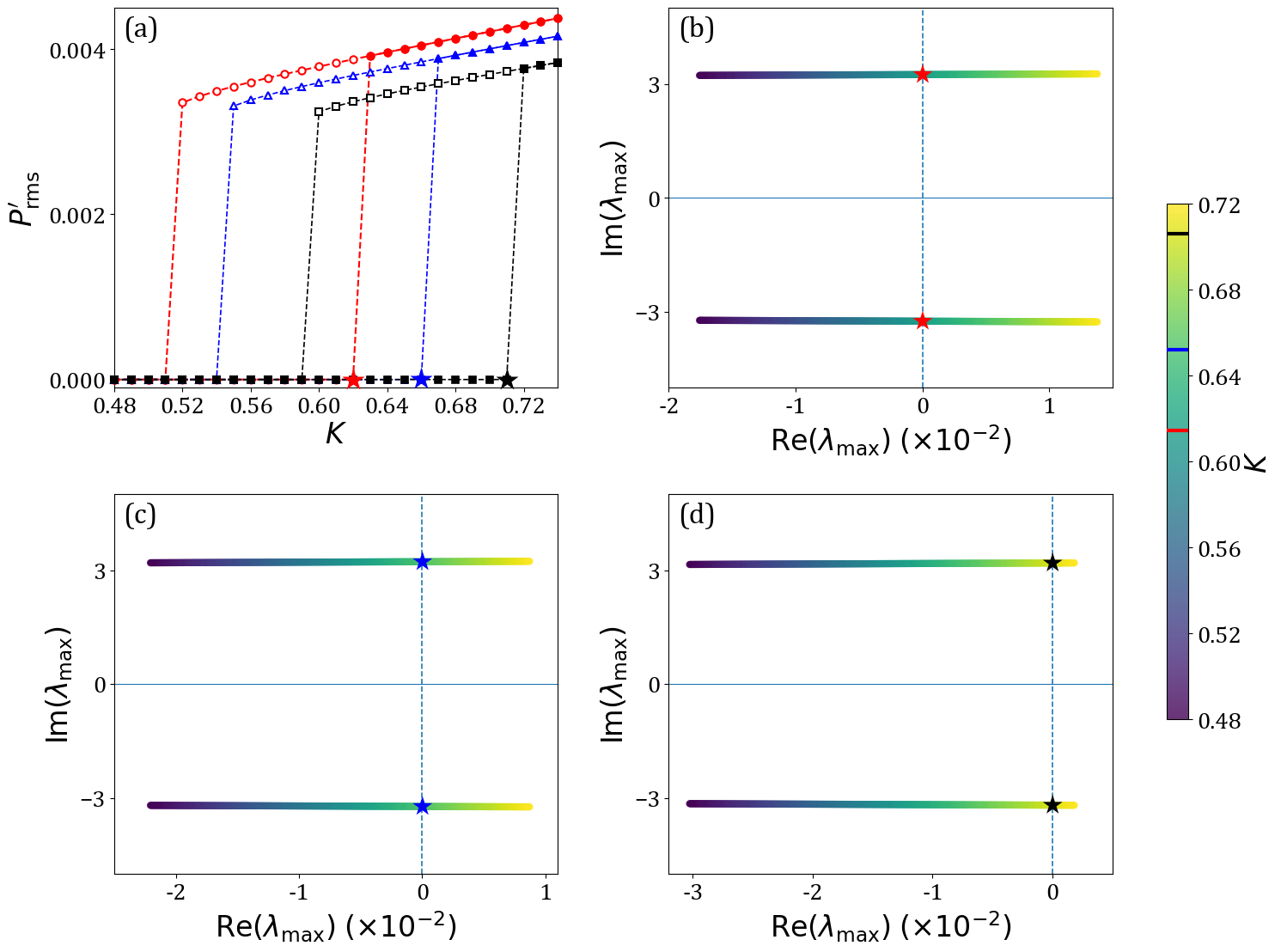}
    \caption{\justifying 
    \textbf{(a)} The variation of $P'_{\mathrm{rms}}$ with heater power $K$ for a single Rijke tube (dashed red line with filled circles for forward simulation and empty circles for backward simulation), time-delay coupled Rijke tubes (dashed blue line with filled triangles for forward simulation and empty triangles for backward simulation), and dissipatively coupled Rijke tubes (dashed black line with filled squares for forward simulation and empty squares for backward simulation). For time-delay coupling, the parameters are fixed at $K_{\tau} = 0.2$ and $\tau = 0.1$, while for dissipative coupling $K_d = 0.4$ and $\omega_B/\omega_A = 0.95$. In both cases, the Hopf and saddle-node bifurcation points shift to higher values of $K$ compared to the single tube. The bistable region exists for $K \in [0.55,0.66]$ for time delay coupling and $K \in [0.60,0.71]$ for dissipative coupling. The stars marked in red, blue, and black colors shows the analytical points for which the occurrence of Hopf point takes place, analytically. \textbf{(b),(c),(d)} Variation of the maximum eigenvalue $\lambda_{\max}$ in the complex eigenvalue plane for the uncoupled Rijke tube, time-delay coupled and dissipatively coupled Rijke tubes as the heater power $K$ is varied shown on the color bar. The red, blue, and black stars correspond to the Hopf point where the $Re(\lambda_{max} = 0)$. The corresponding $K$ values are marked on the color bar with the same color as the star. The theoretical Hopf points are also marked in panel~(a) using the same star colors as in panels~(b)--(d), providing a direct comparison with the numerically observed transition points.
    }
    \label{double_bifur}
\end{figure}
The system is linearly stable when the rightmost characteristic root, $\lambda_{\max}$, satisfies
$\operatorname{Re}(\lambda_{\max})<0$, whereas
$\operatorname{Re}(\lambda_{\max})>0$ indicates a loss of linear
stability. The stability boundary is reached when
$\operatorname{Re}(\lambda_{\max})=0$. In particular, an oscillatory
instability is associated with a complex-conjugate pair of characteristic
roots crossing the imaginary axis as a system parameter is varied.
Thus, at the critical parameter value, the Hopf condition corresponds
to a complex-conjugate pair of characteristic roots,
$\lambda_{H,\pm}$, lying on the imaginary axis, such that
$\operatorname{Re}(\lambda_H)=0$ and
$\operatorname{Im}(\lambda_H)\neq 0$. Hence, the characteristic spectrum
obtained from Eq.~\eqref{eq:characteristic_equation} provides the
criterion for determining the local linear stability of the coupled
thermoacoustic system and for identifying the onset of oscillatory
instability associated with a Hopf bifurcation.

The analytical stability framework above provides a basis for understanding how coupling modifies the transition points as the heater power $K$ is varied, with the analytically predicted Hopf points shown as stars in Fig.~\ref{double_bifur}, consistent with the observed transition points. For the coupled Rijke tubes, the transition characteristics depend on both the coupling parameters and the heater power. For the time-delay coupling, the parameters are fixed at $K_{\tau} = 0.2$ and $\tau = 0.1$, while for the dissipative coupling they are set to $K_d = 0.4$ and $\omega_B/\omega_A = 0.95$. In both cases, the transition points shift to higher values of $K$ compared to the uncoupled system. The Hopf point occurs at $K = 0.66$ in the time-delay coupling and $K = 0.71$ for the dissipative coupling, while the saddle node point appears at $K = 0.55$ and  $K = 0.60$, respectively. These shifts indicate that coupling alters the stability boundaries and delays the onset of self-sustained oscillations.

\end{appendices}

\bibliography{reference}

\end{document}